\documentclass[british,nofootinbib, final]{revtex4-2}
\PassOptionsToPackage{obeyFinal}{todonotes}
\usepackage[T1]{fontenc}
\usepackage[latin9]{inputenc}
\usepackage{geometry}
\usepackage{color}
\usepackage{array}
\usepackage{multirow}
\usepackage{todonotes}
\usepackage{amsmath}
\usepackage{amssymb}
\usepackage{graphicx}
\usepackage{esint}

\makeatletter

\providecommand{\tabularnewline}{\\}

\renewcommand*\arraystretch{2}

\usepackage{color}
\usepackage{marginnote}

\newcommand{\mk}[1]{#1}

\@ifundefined{showcaptionsetup}{}{%
 \PassOptionsToPackage{caption=false}{subfig}}
\usepackage{subfig}
\makeatother

\usepackage{babel}
\begin{document}
\global\long\def\mpl{m_{\mathrm{Pl}}}%

\global\long\def\lpl{l_{\mathrm{Pl}}}%

\global\long\def\eh#1{\hat{\epsilon}_{#1}}%

\global\long\def\ev{\epsilon_{U}}%

\global\long\def\hv{\eta_{U}}%

\global\long\def\hj{\eta_{FV}}%

\global\long\def\xv{\xi_{U}}%

\global\long\def\n{\mathcal{N}}%

\global\long\def\ej#1{\epsilon_{#1}}%

\global\long\def\e{\mathrm{end}}%

\title{Slow-Roll Inflation in the Jordan Frame}
\author{Mindaugas Kar\v{c}iauskas}
\affiliation{Departamento de F\'isica Te\'orica and Instituto de F\'isica de
Part\'iculas y del Cosmos IPARCOS, Universidad Complutense de Madrid,
E-28040 Madrid, Spain}
\author{José Jaime Terente Díaz}
\affiliation{Departamento de F\'isica Te\'orica and Instituto de F\'isica de
Part\'iculas y del Cosmos IPARCOS, Universidad Complutense de Madrid,
E-28040 Madrid, Spain}
\begin{abstract}
Inflation models based on scalar-tensor theories of gravity are formulated
in the Jordan frame but most often analysed in the Einstein frame.
The transformation between the frames is not always desirable. In
this work we formulate slow-roll conditions in the Jordan frame. This
is achieved by comparing different background quantities and approximations
on the same spatial slice in both frames. We use these approximations
to derive simple equations that can be applied to compute inflation
model observables in terms of Jordan frame quantities only. Finally,
we apply some of the results to analyse generalised induced gravity
models and compare them with the latest observations.
\end{abstract}
\maketitle

\section{Introduction}

The use of modified gravity theories to build inflation models is
as old as the paradigm of inflation itself \citep{Starobinsky:1980te}.
In particular, scalar-tensor theories is a very popular stage to construct
such models \citep{La:1989za,McDonald:1990wy,Steinhardt:1990zx,Spokoiny:1984bd,Accetta:1985du,Lucchin:1985ip,Bezrukov:2007ep}.
Initially the work within these theories was concentrated on finding
inflating solutions. It was realised that the scale factor in the
Jordan frame does not have to be accelerating to realise inflation
\citep{Coule:1997cc,Faraoni:2004pi}. This is in contrast to the requirement
in the Einstein frame. The Jordan frame is distinguished by the Planck
length $\lpl$ being time dependent. Therefore, it is not the scale
factor $a\left(t\right)$ itself that is required to be accelerating
but the ratio of the scale factor to the Planck length must be accelerating,
$\left[a\left(t\right)/\lpl\left(t\right)\right]^{\centerdot\centerdot}>0$.\footnote{To make this expression concise the spacetime slicing is assumed in
which the lapse function in the Einstein frame is $\hat{\n}=1$. This
is an opposite from what is used in the rest of this work.}  

Since the introduction of those early models, the precision of cosmological
measurements increased drastically. It is no longer enough to construct
a modified gravity model that solely provides inflation, one needs
to make sure that it satisfies tight observational constraints derived
from the properties of the primordial curvature perturbation.

Given a model of inflation in the Einstein frame, very simple methods
have been developed to compute the properties of the curvature perturbation.
These methods provide relations between the evolution of the homogeneous
mode and the perturbations. In particular, for single field models
of slow-roll inflation, which is the main interest of the current
work, those relations can be written as (see e.g. \citep{Lyth:2009zz})
\begin{eqnarray}
A_{\mathrm{s}} & = & \frac{U}{24\pi^{2}\ev}\,,\label{As}\\
n_{\mathrm{s}}-1 & = & 2\hv-6\ev\,,\label{ns}\\
r & = & 16\ev\,.\label{r}
\end{eqnarray}
The quantities on the L.H.S of the above equations specify spectral
properties of the primordial perturbation: the amplitude and the spectral
index of scalar perturbations as well as the tensor-to-scalar ratio
respectively. On the R.H.S. we have only homogeneous quantities related
to the shape of the potential $U$. They are expressed in terms of
slow-roll parameters defined by
\begin{eqnarray}
\ev & \equiv & \frac{1}{2}\left(\frac{U_{,\varphi}}{U}\right)^{2}\,,\label{eV-Ef}\\
\hv & \equiv & \frac{U_{,\varphi\varphi}}{U}\,,\label{hV-Ef}
\end{eqnarray}
where $\varphi$ is the inflaton with the potential $U\left(\varphi\right)$
and the subindices `,$\varphi$' denote derivatives with respect
to the homogeneous field $\varphi$.

To compare with observations, these quantities are evaluated at the
time when the pivot scale exits the horizon. Typically this happens
in the range of $\hat{N}=45$ to $70$ e-folds before the end of inflation,
depending on the process of reheating \citep{Lyth:2009zz}.\footnote{We use hats to denote geometric quantities in the Einstein frame.}

Instead of using ``potential slow-roll parameters'' $\ev$ and $\hv$
as above we can write analogous relations in terms of the Hubble-flow
functions \citep{Stewart:1993bc,Gong:2001he,Leach:2002ar}. The latter
being defined as
\begin{equation}
\eh 1\equiv-\frac{\mathrm{d}\hat{H}/\mathrm{d}\hat{\tau}}{\hat{H}^{2}}\,,\quad\eh{i+1}\equiv\frac{\mathrm{d}\eh i/\mathrm{d}\hat{\tau}}{\eh i\hat{H}}\,,\label{e1-def}
\end{equation}
where $\hat{\tau}$ is the proper time and $\hat{H}$ is the Hubble
parameter, both to be defined later. 

Observations constraint the first few parameters to be very small
at the time the observable Universe exits the horizon, $\eh 1,\,\eh 2\ll1$
\citep{Planck:2018jri}. In this limit and at the lowest order in
small parameters, the two sets of these parameters can be related
by 
\begin{align}
\eh 1 & \simeq\ev\,,\label{e1-sr}\\
\eh 2 & \simeq4\ev-2\hv\,.\label{e2-sr}
\end{align}
 We can therefore write eqs.~(\ref{As})-(\ref{r}) also in terms
of the Hubble-flow functions. Plugging eqs.~(\ref{e1-sr})-(\ref{e2-sr})
into eqs.~(\ref{As})-(\ref{r}) the lowest order result is\footnote{The higher order expressions can be found, for example, in \citep{Planck:2013jfk}.}
\begin{eqnarray}
A_{\mathrm{s}} & = & \frac{U}{24\pi^{2}\eh 1}\,,\label{As-Hf}\\
n_{\mathrm{s}}-1 & = & -2\eh 1-\eh 2\,,\label{ns-Hf}\\
r & = & 16\eh 1\,.\label{r-Hf}
\end{eqnarray}

The above expressions can be applied to models formulated in the Einstein
frame. However, when dealing with inflation models in the context
of scalar-tensor theories of gravity, they are usually formulated
in the Jordan frame. In that frame, we cannot apply the above results
directly. In many cases this does not present any problems. One can
conformally transform the metric and rewrite the same model in the
Einstein frame, where the gravitational part of the action reduces
to the Einstein-Hilbert form. In this frame the above results can
be readily applied. 

But such a transformation is not always desirable \mk{(see for example
ref.~\citep{Domcke:2017rzu})}. One attractive feature of the Jordan
frame expressions is that typically matter degrees of freedom are
minimally coupled to the gravitational ones. The matter Lagrangian
takes the standard form, which is familiar from quantum field theories
in flat spacetime with constant coupling constants. Once we transform
the action into the Einstein frame, the ``gravitational degrees of
freedom'' get mixed with the matter degrees of freedom and the simple
expressions are lost. Moreover, it is sometimes the case that the
dynamical analysis of the system is much simpler in the Jordan frame
too. 

In all those cases it would be very convenient to have Jordan frame
analogous expressions to eqs.~(\ref{As})-(\ref{r}) and (\ref{As-Hf})-(\ref{r-Hf})
in order to be able to compute inflation observables without ever
transforming the model into the Einstein frame. The goal of this work
is to derive such relations. 

In this paper we consider only the simplest case: scalar-tensor theories
with a single scalar field and the action
\begin{eqnarray}
S & = & \int\mathrm{d}^{4}x\sqrt{-g}\left[F\left(\phi\right)R+\mathcal{L}\left(\phi\right)+\ldots\right]\,,
\end{eqnarray}
where the non-minimal function $F$ is positive and dots represent
matter fields. In the current investigation we assume that matter
fields are negligible and do not affect the dynamics in the regime
we are concerned with. One can, of course, construct models where
such fields determine the evolution of the system significantly \citep{Kuusk:2016rso}.
In addition, such fields do play the crucial role in the process of
reheating. But we are concerned with the dynamics prior to reheating
and assume the matter part of the action is such that the inflationary
dynamics are solely determined by the scalar field $\phi$. Such restrictions
are made in order to limit ourselves to a class of models where the
conformal transformation can be easily performed, so that our results
can be easily checked. We hope to extend the present analysis to more
generic setups in the future.

There have already been a number of works that study cosmological
perturbations in the Jordan frame \mk{\citep{Hwang:1990re,Hwang:1990jh,Hwang:1996xh,Myrzakulov:2015qaa,Elizalde:2018now}}
and study the equivalence between the curvature perturbation in the
two frames \citep{Fakir:1990zi,Makino:1991sg,Gong:2011qe,White:2012ya,Kubota:2011re}.
A frame independent formulation of some late universe processes is
developed in \mk{\citep{Catena:2006bd,Catena:2006gk,Kuusk:2016rso,Racioppi:2021jai}}
and analogous computations related to inflation and its observables
in \citep{Kaiser:1995nv,Prokopec:2013zya,Kuusk:2016rso,Jarv:2016sow}. 

Our approach is different from the mentioned references. Instead of
computing the curvature perturbation in the Jordan frame and analysing
its relation with the analogous quantities in the Einstein frame,
we are only interested in how \emph{homogeneous} quantities, that
are needed to compute inflation observables (such as in eqs.~(\ref{As})-(\ref{r})
and (\ref{As-Hf})-(\ref{r-Hf})), map from \mk{the} Einstein frame
to the Jordan frame under the conformal transformation. By doing so,
we are able to apply observational constraints to models that are
formulated and analysed solely using homogeneous equations in the
Jordan frame. Our method allows for a systematic investigation of
these issues and generalises similar analysis done, for example, in
\mk{ref.~\citep{Garcia-Bellido:1993fsr,Torres:1996fr,Morris:2001ad,Chiba:2008ia,Akin:2020mcr}
and dynamical analysis in ref.~\citep{Jarv:2021qpp}}.

Whenever discussing modified gravity theories of this type, there
is always a question of equivalence of the Einstein and Jordan frames.
Although, some disagreement remains in regards to systems dominated
by the quantum contributions, in regards to our setup, where the classical
homogeneous mode dominates over small quantum fluctuations, the equivalence
between the frames is well established \mk{\citep{Flanagan:2004bz,Faraoni:2006fx,Gong:2011qe}}.\mk{\footnote{See, however, ref.~\citep{Azri:2018gsz} for a differing view.}}

In this work we use geometrical units where $c=\hbar=\mpl=1$, $\mpl=\left(8\pi G\right)^{-1/2}$
and $G$ is the Newton's gravitational constant. We also adopt the
``mostly positive'' signature of the metric.

\section{Conformal Frames and the Conformal Transformation}

We start by adopting a coordinate time $t$ that is being used to
slice the 4-dimensional spacetime into the $t=\mathrm{const}$ space-like
hypersurfaces (the foliation of spacetime) \citep{Gourgoulhon:2007ue}
such that they coincide with the constant energy density ones. As
we are only interested in FLRW spacetimes, these slices are also homogeneous
and isotropic. Otherwise, the choice of $t$ is arbitrary. But once
it is chosen, we keep $t$ fixed.\footnote{Eventually we choose the slicing that simplifies the expressions in
the Jordan frame.}

The proper time $\tau$, on the other hand, depends on the metric,
such that
\begin{equation}
\delta\tau=\n\delta t\,,
\end{equation}
where $\n$ is the lapse function given by
\begin{equation}
\n\equiv\sqrt{-g^{\mu\nu}\nabla_{\mu}t\nabla_{\nu}t}\,,\label{n-def}
\end{equation}
and $\nabla_{\mu}t$ is the gradient of $t$.\footnote{More precisely $\nabla_{\mu}t$ are the coordinate components of the
time-like vector field which is the metric dual to the gradient of
$t$, the latter being a 1-form quantity.} Furthermore, to simplify expressions we can fix spatial coordinates
such that the shift vector vanishes, as is the standard practice.

With this choice of spacetime slicing and threading the flat FLRW
metric can be written as 
\begin{eqnarray}
\mathrm{d}s^{2} & = & -\mathcal{\n}^{2}\mathrm{d}t^{2}+a^{2}\left(t\right)\delta_{ij}\mathrm{d}x^{i}\mathrm{d}x^{j}\,.\label{ds2}
\end{eqnarray}
In the rest of the paper we will use an overdot to denote derivatives
with respect to the coordinate time $t$, for example $\dot{\phi}\equiv\mathrm{d}\phi/\mathrm{d}t$.
As the spacetime slicing is fixed throughout, we use the same notation
for both, Einstein as well as Jordan frame quantities. 

One of the roles the metric plays is to define the units of measure
\citep{Dicke:1961gz}. The transformation of the metric results in
the transformation of those units. A particular transformation, that
is the subject of this work, is the conformal one (or Weyl transformation).
It is often denoted by $\Omega$ and written as \citep{Birrell:1982ix}
\begin{eqnarray}
\hat{g}_{\mu\nu}\left(x\right) & = & \Omega^{2}\left(x\right)g_{\mu\nu}\left(x\right)\,.\label{conf-tr}
\end{eqnarray}
It is obvious from eqs.~(\ref{n-def}) and (\ref{ds2}) that the
conformal transformation rescales the lapse function $\n$ and the
spatial metric. However, this does not imply any change in physical
observables, only in their interpretation \citep{Deruelle:2010ht,Domenech:2016yxd}.
For example, applying the conformal transformation to a FLRW metric
we can map all the dynamical equations into the universe which is
static. This procedure does not change physical observables such as
the redshift. In a static frame the redshift is caused by the time
dependence of particle masses. The relation between the emitted and
observed photon wavelengths is exactly the same in both frames.

The natural application that conformal transformation lends to is
the scalar-tensor theories of gravity \citep{Faraoni:2004pi}. Most
commonly, in these theories one modifies the gravity sector, as compared
to General Relativity, leaving the matter sector untouched (see however
\citep{Deruelle:2010ht} for different models). It is said that in
this form the model is expressed in the Jordan frame. Such a frame
is convenient because the matter sector is minimally coupled to gravity.
However, modifications of the gravity sector renders Newton's gravitational
constant time dependent, which might lead to some counterintuitive
behaviour.

To avoid such a behaviour, and possible mistakes associated with it,
one can apply eq.~(\ref{conf-tr}) to rewrite the action in a way
that the gravity sector of the model is described by the Einstein-Hilbert
action. This is called the Einstein frame, where the Newton's gravitational
constant does not change with time. The price to pay for simplifying
the gravity sector of the action in this way is the complication of
the matter sector. The latter becomes directly coupled to the new
degree of freedom, often in a complicated way. Therefore, the transformation
into the Einstein frame is not always desirable. In those cases one
would like to develop methods of computing the relevant observable
quantities solely within the Jordan frame. In the following sections
we will derive a method to perform such computations.

\section{Inflation in the Einstein Frame\label{sec:Efrm}}

In this section we review the general principles of Einstein frame
single field inflation with a non-canonical kinetic term. Normally
one would canonically normalise the field before doing the analysis.
But we keep the non-canonical function explicit. Additionally, to
simplify the expressions, one would choose such a spacetime slicing
that the proper time coincides with the coordinate time. In practice
that means setting the lapse function to $\hat{\n}=1$. But to aid
our discussion about inflation observables in the Jordan frame, we
do not perform any of these simplifications. The complications introduced
in this section, will pay off in the later ones.

To make the distinction between the frames clearer we use the caret
for Einstein frame \emph{geometric} quantities. For example, the Einstein
frame FLRW metric in eq.~(\ref{ds2}) is written as
\begin{eqnarray}
\mathrm{d}\hat{s}^{2} & = & -\hat{\mathcal{\n}}^{2}\mathrm{d}t^{2}+\hat{a}^{2}\left(t\right)\delta_{ij}\mathrm{d}x^{i}\mathrm{d}x^{j}\,.\label{ds2-Efrm}
\end{eqnarray}

With every constant time hypersurface one can associate the intrinsic,
$^{3}\hat{R}_{\mu\nu}$, as well as the extrinsic, $\hat{K}_{\mu\nu}$,
curvature tensors. The trace of the latter determines the volume expansion
rate \citep{Poisson:2009pwt}
\begin{eqnarray}
\hat{K} & = & -\frac{1}{\delta\hat{\mathcal{V}}}\frac{\mathrm{d}\delta\hat{\mathcal{V}}}{\mathrm{d}\hat{\tau}}\,,
\end{eqnarray}
where $\hat{K}\equiv\hat{K}_{\mu}^{\mu}$, $\hat{\tau}$ is the proper
time and $\delta\hat{\mathcal{V}}\propto\hat{a}^{3}$ is the proper
volume element. But instead of $\hat{K}$ it is conventional to use
the Hubble parameter given by
\begin{eqnarray}
\hat{H} & \equiv & -\frac{1}{3}\hat{K}\,.\label{H-def}
\end{eqnarray}
In terms of the coordinate time $t$ we can write the above expression
as
\begin{eqnarray}
\hat{H} & = & \frac{\dot{\hat{a}}}{\mathcal{\hat{\n}}\hat{a}}\,.\label{H-Efrm}
\end{eqnarray}

Inflation is defined as the period in the history of the Universe
when the expansion rate of the space-like slices is accelerating,
that is
\begin{eqnarray}
\frac{\mathrm{d}^{2}\hat{a}}{\mathrm{d}\hat{\tau}^{2}} & > & 0\,.\label{inf-def}
\end{eqnarray}
We can relate this condition to the time evolution of the Hubble parameter
in eq.~(\ref{H-Efrm}). If we define the first Hubble-flow function
as in eq.~(\ref{e1-def}), the above condition is equivalent to 
\begin{eqnarray}
\eh 1 & < & 1\,.\label{inf-cond}
\end{eqnarray}
That is, the Hubble parameter must be changing slowly. Once this condition
is broken, inflation ends
\begin{eqnarray}
\eh{1\mathrm{end}} & \equiv & 1\,.\label{inf-end}
\end{eqnarray}

Observable scales exit the horizon somewhere between $\hat{N}=45$
and $70$ e-folds before the end of inflation \citep{Lyth:2009zz}.
The current constrains on the Hubble-flow functions at horizon exit
are $\eh{1*}<0.0052$ (95\% CL), $\eh{2*}=0.034\pm0.008$ (68\% CL),
$\eh{3*}=0.13_{-0.45}^{+0.40}$ (95\% CL) \citep{Planck:2018jri},
where $\eh i$ are defined in eq.~(\ref{e1-def}). That is, all these
parameters are much smaller than unity
\begin{eqnarray}
\left|\eh i\right|_{*} & \ll & 1\,,\label{Hfl-small}
\end{eqnarray}
where $i=1,2,3$ and the asterisk denotes the moment when cosmological
scales exit the horizon. We will use this fact later to approximate
many equations.

In the above discussion we considered only geometric quantities, without
any reference to the matter content. Next, we specify the general
action for the single field inflation models, which can be written
as
\begin{eqnarray}
S & = & \int\sqrt{-\hat{g}}\left[\frac{1}{2}\hat{R}-\frac{1}{2}\mathcal{K}\left(\phi\right)\hat{g}^{\mu\nu}\partial_{\mu}\phi\partial_{\nu}\phi-U\left(\phi\right)\right]\,,\label{S-Ef}
\end{eqnarray}
where $\hat{g}\equiv\mathrm{det}\left(\hat{g}_{\mu\nu}\right)$, $\hat{R}$
is the Ricci curvature scalar and $U\left(\phi\right)$ is the potential.
Variation of the action in eq.~(\ref{S-Ef}) with respect to the
field $\phi$ gives the Klein-Gordon equation
\begin{eqnarray}
\phi''+3\hat{H}\phi'+\frac{\mathcal{K}_{,\phi}}{\mathcal{K}}\frac{1}{2}\left(\phi'\right)^{2}+\frac{U_{,\phi}}{\mathcal{K}} & = & 0\,,\label{eom-tau}
\end{eqnarray}
where the primes denote derivatives with respect to the proper time
$\hat{\tau}$. 

The variation of the same action with respect to the metric tensor
leads to the Einstein equation with the energy-momentum tensor given
by
\begin{eqnarray}
T_{\mu\nu} & = & \mathcal{K}\left(\phi\right)\nabla_{\mu}\phi\nabla_{\nu}\phi-g_{\mu\nu}\left[\frac{1}{2}\mathcal{K}\left(\phi\right)\nabla_{\sigma}\phi\nabla^{\sigma}\phi+U\left(\phi\right)\right]\,.
\end{eqnarray}
From the above expression follows that that the energy density and
pressure of the scalar field can be written as
\begin{eqnarray}
\rho=-T_{0}^{0} & = & \frac{1}{2}\mathcal{K}{\phi'}^{2}+U\left(\phi\right)\,,\label{rho}\\
P=\frac{1}{3}T_{i}^{i} & = & \frac{1}{2}\mathcal{K}{\phi'}^{2}-U\left(\phi\right)\,.\label{P}
\end{eqnarray}
Taking the `$00$' and `$ii$' components of the Einstein equation
one arrives at Friedmann equations 
\begin{eqnarray}
\hat{H}^{2} & = & \frac{\frac{1}{2}\mathcal{K}{\phi'}^{2}+U\left(\phi\right)}{3}\,,\label{Fdm-eq}\\
\hat{H}' & = & -\frac{1}{2}\mathcal{K}{\phi'}^{2}\,,\label{cont-eq}
\end{eqnarray}
where we used the expressions for $\rho$ and $P$ given in eqs.~(\ref{rho})
and (\ref{P}) respectively. 

Using the conditions in eq.~(\ref{Hfl-small}) and the equation of
motion in eq.~(\ref{eom-tau}) it is easy to show that eqs.~(\ref{Fdm-eq})
and (\ref{cont-eq}) imply \footnote{The condition $\left|\eh 2\right|\ll1$ implies 
\begin{eqnarray*}
\left|\frac{\phi''}{\phi'\hat{H}}+\frac{1}{2}\frac{\mathcal{K}_{,\phi}}{\mathcal{K}}\frac{\phi'}{\hat{H}}\right| & \ll & 1\,,
\end{eqnarray*}
and eq.~(\ref{eom-tau}) can be written as
\begin{eqnarray*}
-\left(\frac{\phi''}{\hat{H}\phi'}+\frac{\mathcal{K}_{,\phi}\phi'}{2\mathcal{K}\hat{H}}\right) & = & 3+\frac{U_{,\phi}}{\hat{H}\phi'\mathcal{K}}\,.
\end{eqnarray*}
The result in eq.~(\ref{sr2-Ef}) follows from these two relations
above.}
\begin{eqnarray}
\frac{1}{2}\mathcal{K}{\phi'}^{2} & \ll & U\left(\phi\right)\,,\label{sr1-Ef}\\
\left|\frac{\phi''}{\phi'\hat{H}}+\frac{\mathcal{K}_{,\phi}}{2\mathcal{K}}\frac{\phi'}{\hat{H}}\right| & \ll & 3\simeq\left|\frac{U_{,\phi}}{\hat{H}\mathcal{K}\phi'}\right|\,.\label{sr2-Ef}
\end{eqnarray}
One can immediately notice from the last expression that, in contrast
to the case of the canonically normalised field, $\phi''/\left(\phi'\hat{H}\right)$
does not have to be small if $\mathcal{K}\left(\phi\right)$ is a
function that makes the two terms on the L.H.S. in eq.~(\ref{sr2-Ef})
cancel out. 

Applying the above conditions to the Friedman equation (\ref{Fdm-eq})
we find
\begin{eqnarray}
\hat{H}^{2} & \simeq & \frac{U\left(\phi\right)}{3}\,.\label{Hsr}
\end{eqnarray}
Similarly, the condition in eq.~(\ref{sr2-Ef}) applied to eq.~(\ref{eom-tau})
yields a simplified equation of motion 
\begin{eqnarray}
3\hat{H}\mathcal{K}\phi' & \simeq & -U_{,\phi}\,.\label{eom-sr}
\end{eqnarray}
Both eqs.~(\ref{Hsr}) and (\ref{eom-sr}) are called slow-roll equations.

Slow-roll approximation also contains the assumption that the derivative
of the above expression holds \citep{Lyth:2009zz}. This leads to
the expression
\begin{eqnarray}
\frac{\phi''}{\hat{H}\phi'}+\frac{\mathcal{K}_{,\phi}}{2\mathcal{K}}\frac{\phi'}{\hat{H}} & \simeq & \frac{1}{\mathcal{K}}\left[\frac{1}{2}\left(\frac{U_{,\phi}}{U}\right)^{2}+\frac{\mathcal{K}_{,\phi}}{2\mathcal{K}}\frac{U_{,\phi}}{U}-\frac{U_{,\phi\phi}}{U}\right]\,,\label{ddphi-sr}
\end{eqnarray}
where the condition in eq.~(\ref{sr1-Ef}) was applied to the Friedmann
equation (\ref{Fdm-eq}). 

As one can see from eq.~(\ref{sr2-Ef}), slow-roll condition implies
that the R.H.S. of (\ref{ddphi-sr}) is small. This can be conveniently
expressed using slow-roll parameters defined as
\begin{eqnarray}
\ev & \equiv & \frac{1}{2\mathcal{K}}\left(\frac{U_{,\phi}}{U}\right)^{2}\,,\label{eU-def}\\
\hv & \equiv & \frac{1}{\mathcal{K}}\left(\frac{U_{,\phi\phi}}{U}-\frac{\mathcal{K}_{,\phi}}{2\mathcal{K}}\frac{U_{,\phi}}{U}\right)\,.\label{hU-def}
\end{eqnarray}
The above definitions are equivalent to the definitions in eqs.~(\ref{eV-Ef})
and (\ref{hV-Ef}). This can be easily confirmed if we canonically
normalise the field, such that $\mathrm{d}\varphi\equiv\sqrt{\mathcal{K}}\mathrm{d}\phi$.
The normalisation leads to slow-roll parameters as they are written
in those equations. 

The slow-roll relations in eqs.~(\ref{Hsr}), (\ref{eom-sr}) and
the condition in eq.~(\ref{sr1-Ef}) lead to
\begin{eqnarray}
\ev & \ll & 1\,.\label{ev-sr}
\end{eqnarray}
Similarly, applying the condition in eq.~(\ref{sr2-Ef}) to the expression
in eq.~(\ref{ddphi-sr}) we find
\begin{eqnarray}
\left|\hv\right| & \ll & 1\,.\label{hv-sr}
\end{eqnarray}
Notice that the second condition does not necessarily imply $\left|U_{,\phi\phi}\right|/U\ll1$,
as would be the case if the derivatives are taken with respect to
the canonically normalised field. The terms in the parenthesis of
eq.~(\ref{hU-def}) can approximately cancel out, even if each of
them is not small separately.

Using the slow-roll equation of motion in eq.~(\ref{eom-sr}) and
the approximate Hubble parameter in eq.~(\ref{Hsr}) we can readily
show that 
\begin{eqnarray}
\eh 1 & \simeq & \ev\,.\label{e1eU}
\end{eqnarray}

For later use, it is convenient to rewrite all of the above expressions
in terms of the coordinate time $t$. For example, taking $\delta\hat{\tau}=\hat{\n}\delta t$,
the formulas for the Hubble-flow parameters in eq.~(\ref{e1-def})
become
\begin{eqnarray}
\hat{\epsilon}_{1} & = & -\frac{\dot{\hat{H}}}{\hat{\mathcal{\n}}\hat{H}^{2}}\,,\label{e1-N}\\
\hat{\epsilon}_{i+1} & = & \frac{\dot{\hat{\epsilon}}_{i}}{\mathcal{\hat{\n}}\hat{H}\hat{\epsilon}_{i}}\,,\label{Hfl-N}
\end{eqnarray}
the scalar field equation of motion (\ref{eom-tau}) transforms to
\begin{eqnarray}
\ddot{\phi}+\left(3\frac{\dot{\hat{a}}}{\hat{a}}-\frac{\dot{\hat{\n}}}{\hat{\n}}\right)\dot{\phi}+\frac{\mathcal{K}_{,\phi}}{2\mathcal{K}}\dot{\phi}^{2}+\hat{\n}^{2}\frac{U_{,\phi}}{\mathcal{K}} & = & 0\,,\label{eom}
\end{eqnarray}
and the Friedman and continuity equations (\ref{Fdm-eq}) and (\ref{cont-eq})
can be written as 
\begin{eqnarray}
\left(\frac{\dot{\hat{a}}}{\hat{a}}\right)^{2} & = & \frac{\frac{1}{2}\mathcal{K}\dot{\phi}^{2}+\hat{\n}^{2}U\left(\phi\right)}{3}\,,\label{Fdm-eq-t}\\
\left(\frac{\dot{\hat{a}}}{\hat{a}}\right)^{\centerdot} & = & \frac{\dot{\hat{\n}}}{\hat{\n}}\frac{\dot{\hat{a}}}{\hat{a}}-\frac{1}{2}\mathcal{K}\dot{\phi}^{2}\,.\label{cont-eq-t}
\end{eqnarray}
Notice, that $\dot{\hat{a}}/\hat{a}$ is not equal to the Hubble parameter
as the derivative is taken with respect to the coordinate time $t$
and not the proper time $\hat{\tau}$.

Similarly, we can rewrite the slow-roll conditions in eqs.~(\ref{sr1-Ef})
and (\ref{sr2-Ef}) as 
\begin{eqnarray}
\frac{1}{2}\mathcal{K}\left(\phi\right)\dot{\phi}^{2} & \ll & \hat{\n}^{2}U\left(\phi\right)\,,\label{sr1-Ef-t}\\
\frac{1}{2\hat{H}\mathcal{K}\dot{\phi}^{2}}\left|\frac{\mathrm{d}}{\mathrm{d}t}\left(\frac{\mathcal{K}\dot{\phi}^{2}}{\hat{\n}^{2}}\right)\right| & \ll & \frac{3}{\hat{\n}}\simeq\left|\frac{U_{,\phi}}{\hat{H}\mathcal{K}\dot{\phi}}\right|\,.\label{sr2-Ef-t}
\end{eqnarray}
If these conditions are satisfied, the slow-roll approximate dynamical
equations (\ref{Hsr}) and (\ref{eom-sr}) are given by 
\begin{eqnarray}
3\frac{\dot{\hat{a}}}{\hat{a}}\dot{\phi} & \simeq & -\frac{\hat{\n}^{2}U_{,\phi}}{\mathcal{K}}\,,\label{eom1-Ef}\\
\left(\frac{\dot{\hat{a}}}{\hat{a}}\right)^{2} & \simeq & \frac{\hat{\n}^{2}U\left(\phi\right)}{3}\,.\label{H2-sr-Ef}
\end{eqnarray}

The other utility of slow-roll parameters is that they enable us to
write the observable inflation parameters in such a compact form as
in eqs.~(\ref{As})-(\ref{r}). However, to compute numerical values
of those expressions we are missing one more component. Eqs.~(\ref{As})-(\ref{r})
are supposed to be evaluated at the time when observable scales (or
the pivot scale, more precisely) exit the horizon. The exact time
when this happens depends on the specifics of the reheating scenario.
For the purpose of this work, the precise value is not important.
But for concreteness, when comparing our results with observations,
we will take that to be $\hat{N}=50$ to $60$ e-folds before the
end of inflation, in line with the choice of the Planck \mk{} team
\citep{Planck:2018jri}.

The number of e-folds of inflation $\hat{N}$ is defined as 
\begin{eqnarray}
\hat{N}\equiv\ln\frac{\hat{a}_{\mathrm{end}}}{\hat{a}} & = & \intop_{t}^{t_{\mathrm{end}}}\hat{\n}\hat{H}\mathrm{d}t\,,\label{N-def-Ef}
\end{eqnarray}
where values with the label `$\mathrm{end}$' correspond to the
end of inflation, the latter being defined in eq.~(\ref{inf-end}).
Within the slow-roll approximation this formula can be simplified
as
\begin{eqnarray}
\hat{N} & \simeq & \intop_{\phi_{\mathrm{end}}}^{\phi}\sqrt{\frac{\mathcal{K}}{2\ev}}\mathrm{d}\phi\,.\label{N-sr}
\end{eqnarray}

Therefore, the computation of inflation observables (at least for
these simple models) can be summarised as follows. First, using (\ref{N-sr})
compute the value of the inflaton $\phi_{*}$ that corresponds to
a given \mk{number of} e-folds before the end of inflation. Once
$\phi_{*}$ is known it is very easy to compute $U\left(\phi_{*}\right)$,
$\ev\left(\phi_{*}\right)$ and $\hv\left(\phi_{*}\right)$ (and higher
order parameters if needed). Then plugging this result into eqs.~(\ref{As})-(\ref{r})
we get numerical values of inflation observables which can be compared
with observational constraints.

\section{Inflation in Scalar-Tensor Theories}

Having summarised the procedure of computing the observables of slow-roll
inflation in the Einstein frame, we next discuss a method of \mk{performing
the same computation } in the Jordan frame.

\subsection{The Exact Expressions in the Jordan Frame}

Consider a typical action used in the scalar-tensor theories of gravity
\begin{eqnarray}
S & = & \int\mathrm{d}^{4}x\sqrt{-g}\left[\frac{1}{2}F\left(\phi\right)R-\frac{1}{2}g^{\mu\nu}\partial_{\mu}\phi\partial_{\nu}\phi-V\left(\phi\right)\right]\,.\label{S-Jfr}
\end{eqnarray}
This action also includes models where the scalar field $\phi$ has
a non-canonical kinetic term. This can be shown by performing a field
redefinition $\chi\equiv F\left(\phi\right)$. Defining $\omega\left(\chi\right)\equiv F/\left(F_{\phi}\right)^{2}$
leads to
\begin{eqnarray}
S & = & \int\mathrm{d}^{4}x\sqrt{-g}\left[\frac{1}{2}\chi R-\frac{\omega\left(\chi\right)}{2\chi}g^{\mu\nu}\partial_{\mu}\chi\partial_{\nu}\chi-V\left(\chi\right)\right]\,.
\end{eqnarray}
By setting $\omega=\mathrm{\mathrm{const}}$\mk{, which corresponds
to $F\propto\phi^{2}$, and $V\left(\chi\right)=0$} one obtains
the Brans-Dicke action \citep{Fierz:1956zz,Brans:1961sx}. For the
rest of the paper, we will only use the form of the action in eq.~(\ref{S-Jfr}).

As in the previous section we are only interested in the homogeneous,
flat FLRW spacetime. But in contrast to that section \mk{now} we
do fix the spacetime slicing. Constant time hypersufaces are chosen
such that the coordinate time equals the proper time in the Jordan
frame. That is, in this frame we can write the metric as
\begin{eqnarray}
\mathrm{d}s^{2} & = & -\mathrm{d}t^{2}+a^{2}\left(t\right)\delta_{ij}\mathrm{d}x_{i}\mathrm{d}x_{j}\,,\label{ds2-Jfrm}
\end{eqnarray}
where the lapse function $\n=1$.

Next, we can define analogous geometric quantities as in section \ref{sec:Efrm}.
The Hubble parameter is given by the trace of the extrinsic curvature
as in eq.~(\ref{H-def}) (this time without the hats)
\begin{eqnarray}
H & = & \frac{\dot{a}}{a}\,.\label{H-Jfrm}
\end{eqnarray}

The dynamical equations can be derived in two ways. One way is to
vary the action with respect to the field and its derivative. The
Euler equation then leads to the equation of motion for the $\phi$
field. Varying the same action with respect to the metric leads to
the gravitational equations, which can be used to find the Friedman
equations. 

On the other hand, one can arrive at the same result by using the
expressions in section~\ref{sec:Efrm}. To do that, notice that the
action in eq.~(\ref{S-Jfr}) is transformed into the Einstein frame
by plugging $\Omega=\sqrt{F}$ into eq.~(\ref{conf-tr}). This leads
to the action in eq.~(\ref{S-Ef}) where the kinetic function and
the potential are equal to
\begin{eqnarray}
\mathcal{K}\left(\phi\right) & = & \frac{1}{F}+\frac{3}{2}\left(\frac{F_{,\phi}}{F}\right)^{2}\,,\label{K}
\end{eqnarray}
and
\begin{eqnarray}
U\left(\phi\right) & = & \frac{V\left(\phi\right)}{F^{2}}\,,\label{U}
\end{eqnarray}
respectively. As we keep the spacetime slicing fixed even when changing
the frame, the lapse function and the scale factor in the Einstein
frame are functions of $F$
\begin{eqnarray}
\hat{\n} & = & \sqrt{F}\,,\label{N-F}\\
\hat{a} & = & \sqrt{F}a\,,\label{a-F}
\end{eqnarray}
where $a\left(t\right)$ and $\hat{a}\left(t\right)$ are Jordan and
Einstein frame scale factors respectively (see eqs.~(\ref{ds2-Jfrm})
and (\ref{ds2-Efrm})). Plugging these expressions into eq.~(\ref{eom-tau}),
(\ref{Fdm-eq-t}) and (\ref{cont-eq-t}) leads to the equation of
motion, Friedman equation and the continuity equation in terms of
Jordan frame quantities respectively
\begin{eqnarray}
\ddot{\phi}+3H\dot{\phi}+V_{,\phi} & = & 3F_{,\phi}\left(2H^{2}+\dot{H}\right)\,,\label{EoM-Jf}
\end{eqnarray}
and
\begin{eqnarray}
H^{2} & = & \frac{\frac{1}{2}\dot{\phi}^{2}+V-3H\dot{F}}{3F}\,,\label{H2-Jf}\\
\dot{H} & = & -\frac{\dot{\phi}^{2}-H\dot{F}+\ddot{F}}{2F}\,.\label{dH-Jf}
\end{eqnarray}

Note, that due to the forcing term on the R.H.S. of eq.~(\ref{EoM-Jf})
we can no longer assume that starting from initial condition\mk{s}
at rest the field always rolls down towards the minimum of the potential
$V$. In contrast to the Einstein frame intuition, the field can also
climb up the potential. It is straightforward to understand this behaviour,
if we look at the Einstein frame potential in eq.~(\ref{U}). Taking
the derivative with respect to $\phi$ we find
\begin{eqnarray}
U_{,\phi} & = & U\left(\frac{V_{,\phi}}{V}-2\frac{F_{,\phi}}{F}\right)\,.\label{Ugrad}
\end{eqnarray}
The gradient of $U$ determines the direction of the gravitational
force. The Jordan frame potential $V$ does not have the same significance,
and the gravitational force is not uniquely determined by the gradient
of $V$. Depending on the gradient of $F$, the gravitational force
can be directed in the uphill direction of $V$. This is due to the
Newton's gravitational ``constant'' being the function of the field
$\phi$.\footnote{See ref.~\citep{Kodama:2021yrm} for a similar discussion.}

\subsection{Slow-Roll Approximations}

The advantage of computing Jordan frame relations by the above procedure
is that the same substitutions can be used to find unambiguously slow-roll
conditions and equations in this frame. 

To discuss slow-roll inflation we found it to be of great use to introduce
the following set of parameters. First, similarly to the Einstein
frame Hubble-flow parameters, we introduce analogous ones in the Jordan
frame
\begin{equation}
\ej 1\equiv-\frac{\dot{H}}{H^{2}}\,,\quad\ej{i+1}\equiv\frac{\dot{\ej i}}{\ej iH}\,.\label{e1-Jfrm}
\end{equation}
The time evolution of the non-minimal function $F$ can be also conveniently
parametrised by the following hierarchy of parameters 
\begin{equation}
\theta_{1}\equiv\frac{\dot{F}}{2HF}\,,\quad\theta_{i+1}\equiv\frac{\dot{\theta}_{i}}{H\theta_{i}}\,.\label{theta1-def}
\end{equation}
\mk{In principle, these two sets of parameters are sufficient to
describe the system. However, we find that equations are more compact
and slow-roll approximations are made clearer if we trade Hubble flow
parameters $\ej i$ in favour of $\gamma_{i}$ defined as}
\begin{eqnarray}
\gamma^{2} & \equiv & \frac{\mathcal{K}\dot{\phi}^{2}}{2H^{2}}\,,\quad\gamma_{i+1}\equiv\frac{\dot{\gamma}_{i}}{H\gamma_{i}}\,,\label{gm-def}
\end{eqnarray}
where $\gamma_{1}\equiv\gamma$ and $\mathcal{K}$ is defined in
eq.~(\ref{K}). \mk{One can readily demonstrate the relation between
the three sets to be
\begin{eqnarray}
\gamma^{2} & = & \left(\ej 1+\theta_{1}\right)\left(1+\theta_{1}\right)-\theta_{1}\theta_{2}\,,\label{gamma2-theta}
\end{eqnarray}
at the lowest order. Taking time derivatives would give relations
between higher order parameters. For convenience we call these three
sets as Jordan frame flow parameters. As demonstrated bellow, $\ej 1$,
$\theta_{1}$ and $\gamma^{2}$} all have to be small in slow-roll.

Let us first consider the Hubble flow parameter in eq.~(\ref{e1-N})
and the condition for inflation in eq.~(\ref{inf-cond}). Plugging
first eqs.~(\ref{N-F}) and (\ref{a-F}) into eq.~(\ref{H-Efrm})
we find
\begin{eqnarray}
\hat{H} & = & \frac{H}{\sqrt{F}}\left(1+\theta_{1}\right)\,.\label{H-EJfrms}
\end{eqnarray}
Taking the time derivative on both sides and plugging the result into
eq.~(\ref{e1-N}) we can express $\eh 1$ in terms of the Jordan
frame quantities
\begin{eqnarray}
\eh 1 & = & \frac{\ej 1+\theta_{1}}{1+\theta_{1}}-\frac{\theta_{1}\theta_{2}}{\left(1+\theta_{1}\right)^{2}}\,,\label{e1-EJfrms}
\end{eqnarray}
Imposing the bound in eq.~(\ref{inf-cond}) leads to the condition
for inflation in the Jordan frame
\begin{eqnarray}
\ej 1 & < & 1+\frac{\theta_{1}\theta_{2}}{1+\theta_{1}}\,.\label{inf-cond-Jfrm}
\end{eqnarray}
This result demonstrates explicitly a known fact: the universe can
be inflating even if the first Hubble-flow function is larger than
unity in the Jordan frame. 

To make this point even stronger we can find the relation between
the second time derivative of the scale factor in the two frames 
\begin{eqnarray}
\hat{a}'' & = & \frac{1}{\sqrt{F}}\left[\ddot{a}\left(1+\theta_{1}\right)+aH^{2}\theta_{1}\theta_{2}\right]\,.
\end{eqnarray}
Applying the definition of inflation in eq.~(\ref{inf-def}) to the
above expression we find 
\begin{eqnarray}
\frac{\ddot{a}}{aH^{2}} & > & -\frac{\theta_{1}\theta_{2}}{1+\theta_{1}}\,.\label{a-infl}
\end{eqnarray}
In the Einstein frame the acceleration of the scale factor is the
requirement for inflation. But as can be seen from the above condition,
the Jordan frame scale factor does not have to be accelerating in
order to have inflation \citep{Coule:1997cc,Faraoni:2004pi}, as was
already mentioned in the Introduction. In other words, even if the
acceleration of the Einstein frame scale factor on a given spacetime
slice is positive, the acceleration of the Jordan frame scale factor
on that same slice can be negative. 

The condition in eq.~(\ref{inf-cond-Jfrm}) is the necessary condition
for the universe to be inflating. However, when cosmological scales
exit the horizon, observations impose much stronger constraints. Generically
one must fulfil the slow-roll conditions in eqs.~(\ref{sr1-Ef-t})
and (\ref{sr2-Ef-t}). Plugging in eqs.~(\ref{K}), (\ref{U}), (\ref{N-F})
and (\ref{a-F}) into eq.~(\ref{sr1-Ef-t}) leads to the first slow-roll
relation in the Jordan frame  
\begin{eqnarray}
F\mathcal{K}\dot{\phi}^{2} & \ll & 2V\,,\label{sr1-Jrm-v1}
\end{eqnarray}
or equivalently 
\begin{eqnarray}
\gamma^{2} & \ll & \frac{V}{H^{2}F}\,.\label{sr1-Jrm-v3}
\end{eqnarray}
Plugging, next, those same equations into eq.~(\ref{sr2-Ef-t}) we
find the second slow-roll relation 
\begin{eqnarray}
\frac{1}{2}\left|\frac{\left(\mathcal{K}\dot{\phi}^{2}\right)^{\bullet}}{\mathcal{K}\dot{\phi}^{2}}-\frac{\dot{F}}{F}\right| & \ll & 3\left(\frac{\dot{F}}{2F}+H\right)\simeq\frac{V}{F\mathcal{K}\dot{\phi}^{2}}\left|\frac{\dot{V}}{V}-2\frac{\dot{F}}{F}\right|\,.\label{sr2-Jrm-v1}
\end{eqnarray}
In terms of the $\ej 1$, $\theta_{1}$ and $\gamma^{2}$ parameters,
this condition can also be written as
\begin{eqnarray}
\left|\gamma_{2}-\ej 1-\theta_{1}\right| & \ll & 3\left(1+\theta_{1}\right)\simeq\frac{V}{2H^{2}F\gamma^{2}}\left|\frac{\dot{V}}{HV}-4\theta_{1}\right|\,.\label{sr2-Jrm-v2}
\end{eqnarray}

The above relations are obtained by the direct mapping of the Einstein
frame slow-roll conditions into the Jordan frame. We proceed next
to investigate their implications for the Jordan frame dynamical equations. 

First, we rewrite Eq.~(\ref{sr1-Jrm-v1}) as $\dot{\phi}^{2}+\frac{3}{2}\frac{\dot{F}^{2}}{F}\ll2V$.
As both terms on the L.H.S. of this expression are positive, the condition
has to be satisfied for each of the terms separately. That is
\begin{eqnarray}
\frac{1}{2}\dot{\phi}^{2} & \ll & V\,,\label{sr1-Jrm-v2-t1}\\
\frac{1}{2}\theta_{1}^{2} & \ll & \frac{V}{3H^{2}F}\,.\label{sr1-Jrm-v2-t2}
\end{eqnarray}
We can next rewrite the Friedmann equation (\ref{H2-Jf}) as
\begin{eqnarray}
1+2\theta_{1} & = & \frac{V+\frac{1}{2}\dot{\phi}^{2}}{3FH^{2}}\,.
\end{eqnarray}
The second term on the R.H.S. of this expression can be dropped due
to the slow-roll condition in eq.~(\ref{sr1-Jrm-v2-t1}). And the
condition in eq.~(\ref{sr1-Jrm-v2-t2}) leads to
\begin{eqnarray}
\frac{1}{2}\theta_{1}^{2} & \ll & 1+2\theta_{1}\simeq\frac{V}{3FH^{2}}\,.
\end{eqnarray}
The inequality can be satisfied only in the case of
\begin{eqnarray}
\left|\theta_{1}\right| & \ll & 1\,.\label{th-small}
\end{eqnarray}
Therefore, the slow-roll approximated Friedmann equation in the Jordan
frame is given by  
\begin{eqnarray}
H^{2} & \simeq & \frac{V}{3F}\,.\label{sr-H-Jfrm}
\end{eqnarray}

To derive the slow-roll equation of motion we make use of its expression
in the Einstein frame in eq.~(\ref{eom1-Ef}) and use the same substitutions
as above. This leads to  
\begin{equation}
3H\dot{\phi}\simeq\frac{2VF_{,\phi}-FV_{,\phi}}{F^{2}\mathcal{K}\left(1+\theta_{1}\right)}\,,\label{sr-eom-Jfrm-th1}
\end{equation}
which can be simplified after imposing the slow-roll condition in
eq.~(\ref{th-small}): 
\begin{eqnarray}
3H\dot{\phi} & \simeq & \frac{2VF_{,\phi}-FV_{,\phi}}{F^{2}\mathcal{K}}\,.\label{sr-eom-Jfrm}
\end{eqnarray}
\mk{It is worth noticing here, as is it pointed out in ref.~\citep{Akin:2020mcr},
that one can find a very similar slow-roll expression used in the
literature: $3H\dot{\phi}\simeq\left(2VF_{,\phi}-FV_{,\phi}\right)/F$
(see for example refs.~\citep{Garcia-Bellido:1993fsr,Morris:2001ad,Chiba:2008ia}).
It is obtained by applying, what is sometimes called, ``generalised
slow-roll'' approximation \citep{Garcia-Bellido:1993fsr,Faraoni:2000nt,Akin:2020mcr}.
Comparing with eq.~(\ref{sr-eom-Jfrm}) we can see that this expression
is equivalent to neglecting $F_{,\phi}^{2}/F$ term. But generally
neglecting this term is not justified. As we show in the section~\ref{sec:genIGr},
the generalised induced gravity model is compatible with observations
only in the region where precisely $F_{,\phi}^{2}\gg F$. Moreover,
by performing computer simulations the authors of ref.~\citep{Jarv:2021qpp}
demonstrate that indeed eq.~(\ref{sr-eom-Jfrm}) is the slow-roll
attractor. In ref.~\citep{Akin:2020mcr} one can find an analysis
of some implications resulting from the two different expressions.}

Having shown that slow-roll inflation requires the time derivative
of $F$ to be small, we can do the same for other functions. First,
we make use of the R.H.S. relation of the slow-roll expression in
eq.~(\ref{sr2-Jrm-v2}). As $\theta_{1}$ is slow-roll suppressed
and  $V/H^{2}F\gamma^{2}\gg1$, according to eq.~(\ref{sr1-Jrm-v3}),
the relation  $3\simeq\left(V/2H^{2}F\gamma^{2}\right)\left|\dot{V}/HV-4\theta_{1}\right|$
(c.f. eq.~(\ref{sr2-Jrm-v2})) leads to 
\begin{eqnarray}
\frac{\dot{V}}{HV} & \ll & 1\,.\label{dV-sr}
\end{eqnarray}
Another set of slow-roll conditions can be found from the L.H.S. of
eq.~(\ref{sr2-Jrm-v2}). Neglecting again $\theta_{1}$ one finds
\begin{eqnarray}
\left|\gamma_{2}-\ej 1\right| & \ll & 1\,.\label{gamma-e1}
\end{eqnarray}
By itself, this slow-roll condition does not require either $\gamma_{2}$
or $\ej 1$ to be small if they are tuned to cancel out. As we will
see shortly, if both of these terms are larger than 1, such a cancellation
also requires one more cancellation with a large $\theta_{2}$ value.
To stay generic, we assume no such cancellations and write
\begin{eqnarray}
\left|\gamma_{2}\right| & \ll & 1\,,\label{g2-ll-1}\\
\left|\ej 1\right| & \ll & 1\,.\label{e1-ll-1}
\end{eqnarray}

To derive other implications of the slow-roll condition we can use
eqs.~(\ref{cont-eq}) and (\ref{H-EJfrms}) to write 
\begin{eqnarray}
\eh 1 & = & \frac{\gamma^{2}}{\left(1+\theta_{1}\right)^{2}}\,.
\end{eqnarray}
Therefore, applying slow-roll conditions in eqs.~(\ref{th-small})
and (\ref{Hfl-small}) we find
\begin{eqnarray}
\frac{\dot{\phi}^{2}}{2H^{2}F} & \ll & 1\,,\label{dphi}
\end{eqnarray}
and
\begin{eqnarray}
\eh 1 & \simeq & \gamma^{2}\,.\label{e1-EJfrms-v2}
\end{eqnarray}

Higher order smallness parameters can be derived by taking higher
derivatives of eq.~(\ref{e1-EJfrms-v2}). For example, the second
Hubble-flow parameter in the Einstein frame can be related to the
Jordan frame flow parameters by
\begin{eqnarray}
\eh 2 & = & \frac{2}{1+\theta_{1}}\left(\gamma_{2}-\frac{\theta_{1}\theta_{2}}{1+\theta_{1}}\right)\,.
\end{eqnarray}
Up to the lowest order in $\theta_{1}$, this expression can be simplified
as
\begin{eqnarray}
\eh 2 & \simeq & 2\left(\gamma_{2}-\theta_{1}\theta_{2}\right)\,.\label{e2-EJfrms}
\end{eqnarray}
 The above result makes it clear that the slow-roll condition $\eh 2\ll1$
can be satisfied even for large values of $\left|\gamma_{2}\right|$
and $\left|\theta_{1}\theta_{2}\right|$, if they \mk{approximately}
cancel out. Paired with eq.~(\ref{gamma-e1}) this leads to $\gamma_{2}\simeq\ej 1\simeq\theta_{1}\theta_{2}$.
However, we do not analyse this case any further, as mentioned above,
and generically write the condition as
\begin{eqnarray}
\left|\theta_{1}\theta_{2}\right| & \ll & 1\,,\label{th2-constr}
\end{eqnarray}
which \mk{still permits $\left|\theta_{2}\right|\sim1$ for small
$\left|\theta_{1}\right|$. Such large $\left|\theta_{2}\right|$
values imply $\left|\ddot{F}\right|\sim H\left|\dot{F}\right|$, which
could provide a counter example to the approximation $\left|\ddot{F}\right|\ll H\left|\dot{F}\right|$,
which is often used in the literature.}

\subsection{Inflation Observables in the Jordan Frame}

In the Einstein frame we can relate the Hubble-flow parameters with
spectral properties of the primordial curvature perturbation as shown
in eqs.~(\ref{As-Hf})-(\ref{r-Hf}). Having derived the relation
between Einstein frame Hubble-flow parameters and analogous quantities
in the Jordan frame in eqs.~(\ref{e1-EJfrms-v2}) and (\ref{e2-EJfrms}),
we can relate them with the spectral properties of the primordial
curvature perturbation as 
\begin{eqnarray}
A_{\mathrm{s}} & \simeq & \frac{V}{24\pi^{2}F^{2}\gamma^{2}}\,,\label{As-Jf}\\
n_{\mathrm{s}}-1 & \simeq & -2\left(\gamma^{2}+\gamma_{2}-\theta_{1}\theta_{2}\right)\,,\label{ns-Jf}\\
r & \simeq & 16\gamma^{2}\,.\label{r-Jf}
\end{eqnarray}
Instead of using the $\gamma$ parameter, we can also express the
above relations in terms of the Jordan frame Hubble flow parameters
defined in eq.~(\ref{e1-Jfrm}). \mk{To do that we first apply slow-roll
approximation to eq.~(\ref{gamma2-theta}), finding
\begin{eqnarray}
\gamma^{2} & \simeq & \ej 1+\theta_{1}-\theta_{1}\theta_{2}\,,\label{gamma2-theta-apx}
\end{eqnarray}
where} we kept the last term because $\left|\theta_{2}\right|$ is
allowed to be of order 1 (see eq.~(\ref{th2-constr})). Taking the
derivative of eq.~(\ref{gamma2-theta-apx}) and plugging the result
into eqs.~(\ref{As-Jf})-(\ref{r-Jf}) we find
\begin{eqnarray}
A_{\mathrm{s}} & \simeq & \frac{V}{24\pi^{2}F^{2}\left(\ej 1+\theta_{1}-\theta_{1}\theta_{2}\right)}\,,\label{As-ev}\\
n_{\mathrm{s}}-1 & \simeq & -2\left(\ej 1+\theta_{1}-\theta_{1}\theta_{2}+\frac{\ej 2\ej 1+\theta_{1}\theta_{2}\left(1-\theta_{2}-\theta_{3}\right)}{2\left(\ej 1+\theta_{1}-\theta_{1}\theta_{2}\right)}\right)\,,\\
r & \simeq & 16\left(\ej 1+\theta_{1}-\theta_{1}\theta_{2}\right)\,.\label{r-ev}
\end{eqnarray}
\mk{The above expressions are consistent with eqs.~(20) in ref.~\citep{Akin:2020mcr}.
But we do not assume $\left|\theta_{2}\right|,\,\left|\theta_{3}\right|\ll1$.
These parameters are allowed to be of order 1 by slow-roll approximation
as it is argued bellow eq.~(\ref{th2-constr}).}

The above equations relate the observable parameters of slow-roll
inflation with Jordan frame quantities. However, these equations are
not sufficient to compare them with observations. We need to specify
the moment at which they must be evaluated. Within the single field
slow-roll models of inflation it is sufficiently accurate to evaluate
the above expressions at the moment when the pivot scale exits the
horizon. In the Einstein frame, this is assumed to happen $\hat{N}_{*}$
e-folds before the end of inflation. The Planck satellite team considers
$\hat{N}_{*}=50$ to $60$. However, the Jordan frame number of e-folds
$N$ does not necessarily correspond to the same numerical value as
$\hat{N}$ \citep{Karam:2017zno,Lerner:2009na,Racioppi:2021jai}.

We define the number of e-folds $N$ in the Jordan frame analogously
to its counterpart in the Einstein frame, which is given in eq.~(\ref{N-def-Ef})
\begin{eqnarray}
N\equiv\ln\frac{a_{\mathrm{end}}}{a} & = & \intop_{t}^{t_{\e}}H\mathrm{d}t\,.\label{N-def-Jf}
\end{eqnarray}
Remember, since we keep the spacetime slicing fixed, the initial slice
at $t$ coincides in the Einstein as well as Jordan frames. The same
can be said about the $t_{\e}$ slice. In other words, the limits
of integration are the same in both frames. Plugging eqs.(\ref{N-F}),(\ref{H-EJfrms})
into eq.~(\ref{N-def-Ef}) and using eq.~(\ref{N-def-Jf}) we get
\begin{eqnarray}
\mathrm{d}\hat{N} & = & \left(1+\theta_{1}\right)\mathrm{d}N\,.\label{dhatN-dN}
\end{eqnarray}
When cosmological scales exit the horizon, $\left|\theta_{1}\right|\ll1$
according to eq.~(\ref{th-small}), making the difference between
$\mathrm{d}\hat{N}$ and $\mathrm{d}N$ slow-roll suppressed. But
$N$ is an integral quantity. Therefore, depending on the behaviour
of $\left|\theta_{1}\right|$ throughout inflation, the total e-fold
shift number between the Einstein and Jordan frames can become observationally
relevant (see the models in section~\ref{subsec:Chaotic-Type}, for
example). We can readily compute this shift number $\int\theta_{1}\mathrm{d}N$.
Using the definition of $\theta_{1}$ in eq.~(\ref{theta1-def})
and the definition of e-fold numbers in eq.~(\ref{N-def-Jf}) gives
 
\begin{eqnarray}
\hat{N} & = & N+\frac{1}{2}\ln\frac{F_{\e}}{F}\,.\label{thN-SR-Jfrm}
\end{eqnarray}

Although in this work we are only concerned about homogeneous quantities
(apart from the numerical simulations in section~\ref{sec:numerical}),
we would also like to comment here about the horizon crossing time
(see also \citep{Racioppi:2021jai}). In the Einstein frame, a mode
of perturbation is said to leave the horizon when its wavenumber is
$k=\hat{a}\hat{H}$ (see e.g. \citep{Lyth:2009zz} for details). Using
the transformations in eqs.~(\ref{H-EJfrms}) and (\ref{a-F}) we
can easily find that this corresponds to
\begin{equation}
\hat{a}\hat{H}=aH\left(1+\theta_{1}\right)\,.
\end{equation}
As we can see, for cosmological scales, where slow-roll conditions
are meant to be applicable, the error of the often used choice $k=aH$
is of the order of slow-roll (see eq.~(\ref{th-small})).

\subsection{Slow-Roll Parameters \label{subsec:SR-Prm}}

Equations (\ref{As-Jf})-(\ref{r-Jf}) are perfectly valid equations
to compute inflation observables in the Jordan frame. They can be
considered as analogous equations to the expressions in terms of the
Hubble-flow functions in the Einstein frame as in eqs.~(\ref{As-Hf})-(\ref{r-Hf}).
However, the computational benefit of the slow-roll attractor is that
time derivatives become unique functions of the field value, provided
by slow-roll equation of motion. Therefore, this effectively reduces
the dynamical degrees of freedom.

In the Einstein frame, this computational benefit is utilised by writing
the inflation observables as functions of the field, as in eqs.~(\ref{As})-(\ref{r}).
We can also do the same in the Jordan frame. 

First, we can use the relation between Einstein frame Hubble-flow
parameters and the Jordan frame parameters in eqs.~(\ref{e1-EJfrms-v2})
and (\ref{e2-EJfrms}). Plugging these into eqs.~(\ref{e1-sr}) and
(\ref{e2-sr}) we find

\begin{eqnarray}
\ev & \simeq\eh 1\simeq & \gamma^{2}\,,\label{e1-gamma}\\
4\ev-2\hv & \simeq\eh 2\simeq & 2\left(\gamma_{2}-\theta_{1}\theta_{2}\right)\,.\label{e2-gamma}
\end{eqnarray}

The consistency of the above result can be checked directly using
eqs.~(\ref{eU-def}) and (\ref{hU-def}). Plugging in the expression
for $U$ in eq.~(\ref{U}) we arrive at 
\begin{eqnarray}
\ev & = & \frac{1}{2\mathcal{K}}\left(\frac{V_{,\phi}}{V}-2\frac{F_{,\phi}}{F}\right)^{2}\,.\label{eu-gamma}
\end{eqnarray}
Using Jordan frame slow-roll equation of motion (\ref{sr-eom-Jfrm}),
slow-roll Friedmann equation (\ref{sr-H-Jfrm}) and the definition
of $\gamma$ in eq.~(\ref{gm-def}) we arrive at (\ref{e1-gamma}).

As mentioned above, slow-roll approximation also contains the assumption
that the time derivative of the slow-roll equation of motion in eq.~(\ref{sr-eom-Jfrm-th1})
holds. This leads to
\begin{eqnarray}
\gamma_{2}-\theta_{1}\theta_{2} & \simeq & \hj\,,\label{dEoM-Jfrm}
\end{eqnarray}
where we defined
\begin{eqnarray}
\hj\equiv2\ev-\hv & = & \frac{1}{\mathcal{K}}\left[2\frac{F_{,\phi\phi}}{F}-\frac{V_{,\phi\phi}}{V}-2\frac{F_{,\phi}^{2}}{F^{2}}+\frac{V_{,\phi}^{2}}{V^{2}}+\frac{\mathcal{K}_{,\phi}}{2\mathcal{K}}\left(\frac{V_{,\phi}}{V}-2\frac{F_{,\phi}}{F}\right)\right]\,,\label{hj-def}
\end{eqnarray}
which, again, could be derived by directly plugging in the expression
for $U$ into the definition of $\hv$ in eq.~(\ref{hU-def}). Notice,
that we did not use eq.~(\ref{sr-eom-Jfrm}) to derive the above
result, because $\theta_{2}$ does not necessarily have to be small.
That is, $\theta_{1}\theta_{2}$ is allowed to be of order slow-roll
and not slow-roll squared.

According to the slow-roll condition in eq.~(\ref{sr2-Jrm-v2}) the
L.H.S. of the above result has to be small. Hence the R.H.S. too.
It follows then, that we can write the second slow-roll constraint
as
\begin{eqnarray}
\left|\hj\right| & \ll & 1\,.\label{sr-hj}
\end{eqnarray}
This can be equally well confirmed by eq.~(\ref{e2-gamma}): $2\eta_{FV}\simeq\eh 2\ll1$.

Plugging eqs.~(\ref{e1-gamma}) and (\ref{dEoM-Jfrm}) into (\ref{As})-(\ref{r})
we arrive at slow-roll expressions which relate inflation observables
with functions $F$ and $V$
\begin{eqnarray}
A_{\mathrm{s}} & \simeq & \frac{V}{24\pi^{2}F^{2}\ev}\,,\label{As-FV}\\
n_{\mathrm{s}}-1 & \simeq & -2\left(\ev+\hj\right)\,,\label{ns-FV}\\
r & \simeq & 16\ev\,,\label{r-FV}
\end{eqnarray}
where $\ev$ is meant to be evaluated using eq.~(\ref{eu-gamma}).

We can also conveniently slow-roll approximate the number of e-folds
of inflation in the Jordan frame. Using the definition of $N$ in
eq.~(\ref{N-def-Jf}) and slow-roll equation of motion in eq.~(\ref{sr-eom-Jfrm})
we arrive at  
\begin{eqnarray}
N & \simeq & \intop_{\phi}^{\phi_{\mathrm{end}}}\frac{\mathcal{K}}{2\frac{F_{,\phi}}{F}-\frac{V_{,\phi}}{V}}\mathrm{d}\phi\,.\label{N-SR-Jfrm-2}
\end{eqnarray}

So far, we have considered a generic expression for $\mathcal{K}$.
But the definition of $\mathcal{K}$ in eq.~(\ref{K}) contains two
positive terms. These terms remain comparable throughout inflation
only for a specific form of $F$, namely
\begin{eqnarray}
F & = & \left(\alpha\phi+\beta\right)^{2}\,,\label{Feq}
\end{eqnarray}
with $\alpha=\mathcal{O}\left(1\right)$. In the case of $\alpha=1/\sqrt{6}$,
both of those terms are equal. More generically, however, one of
them should be dominant. In those cases the expressions can be somewhat
simplified. 

Consider first the case $F_{,\phi}^{2}\ll F$. That is, the approximate
expression for $\mathcal{K}$ is given by
\begin{eqnarray}
\mathcal{K} & \simeq & \frac{1}{F}\,.
\end{eqnarray}
In this case the slow-roll equation of motion (\ref{sr-eom-Jfrm})
can be approximated as 
\begin{eqnarray}
\frac{\dot{\phi}}{H} & \simeq & 2F_{,\phi}-F\frac{V_{,\phi}}{V}\,,\label{srEoM-dF2-small}
\end{eqnarray}
and slow-roll parameter $\ev$ 
\begin{eqnarray}
\ev & \simeq & \frac{F_{,\phi}^{2}}{2F}\left(2-\frac{F}{F_{,\phi}}\frac{V_{,\phi}}{V}\right)^{2}\,.\label{eU-dF2-small}
\end{eqnarray}
The smallness of $\ev$ implies
\begin{eqnarray}
\frac{1}{2}\left(\frac{V_{,\phi}}{V}\right)^{2} & \ll & \frac{1}{F}\,.\label{Vlim-dF2-small}
\end{eqnarray}
The second slow-roll parameter $\eta_{FV}$ (defined in eq.~(\ref{hj-def}))
in this approximations can be simplified as 
\begin{eqnarray}
\hj & \simeq & 2F_{,\phi\phi}+F\left(\frac{V_{,\phi}^{2}}{V^{2}}-\frac{V_{,\phi\phi}}{V}-\frac{F_{,\phi}^{2}}{F^{2}}-\frac{F_{,\phi}}{2F}\frac{V_{,\phi}}{V}\right)\,,\label{hU-dF2-small}
\end{eqnarray}
where we also applied $\left|F_{,\phi\phi}\right|\ll1$. Finally,
the number of e-folds in this regime can be expressed as
\begin{eqnarray}
N & \simeq & \intop_{\phi_{\mathrm{end}}}^{\phi}\frac{1}{\frac{V_{,\phi}}{V}-2\frac{F_{,\phi}}{F}}\frac{\mathrm{d}\phi}{F}\,.\label{eq:N-dF2-small}
\end{eqnarray}

In the opposite case $F_{,\phi}^{2}\gg F$ and
\begin{eqnarray}
\mathcal{K} & \simeq & \frac{3}{2}\frac{F_{,\phi}^{2}}{F^{2}}\,.
\end{eqnarray}
The slow-roll equation of motion can then be approximated as
\begin{eqnarray}
\frac{\dot{\phi}}{H} & \simeq & \frac{2}{3}\frac{F}{F_{,\phi}}\left(2-\frac{F}{F_{,\phi}}\frac{V_{,\phi}}{V}\right)\,,\label{srEoM-dF2-large}
\end{eqnarray}
while the slow-roll parameter $\ev$ is given by
\begin{eqnarray}
\ev & \simeq & \frac{1}{3}\left(2-\frac{F}{F_{,\phi}}\frac{V_{,\phi}}{V}\right)^{2}\,.\label{eU-dF2_large}
\end{eqnarray}
We can see that in this regime slow-roll condition $\ev\ll1$ requires
the cancellation
\begin{eqnarray}
\frac{V_{,\phi}}{V} & \simeq & 2\frac{F_{,\phi}}{F}\,,\label{dVV=00003DdFF}
\end{eqnarray}
up to slow-roll precision. 

The second slow-roll parameter $\eta_{FV}$ can be simplified as
\begin{eqnarray}
\hj & \simeq & \frac{2F^{2}}{3F_{,\phi}^{2}}\left[\frac{V_{,\phi}^{2}}{V^{2}}-\frac{V_{,\phi\phi}}{V}+\left(F_{,\phi\phi}-\frac{F_{,\phi}^{2}}{F}\right)\frac{V_{,\phi}}{F_{,\phi}V}\right]\,.\label{hU-dF2_large}
\end{eqnarray}
Finally, the number of e-folds of inflation in the Jordan frame
is given by 
\begin{eqnarray}
N & \simeq & \frac{3}{2}\intop_{\phi_{\mathrm{end}}}^{\phi}\frac{F_{,\phi}^{2}/F}{\frac{V_{,\phi}}{V}-2\frac{F_{,\phi}}{F}}\frac{\mathrm{d}\phi}{F}\,.\label{N-dF2_large}
\end{eqnarray}

We can also study separately the case $F_{,\phi}^{2}\simeq F$, which
leads to eq.~(\ref{Feq}). In this case 
\begin{eqnarray}
\ev & \simeq & \mathcal{O}\left(1\right)\left(2-\frac{F}{F_{,\phi}}\frac{V_{,\phi}}{V}\right)^{2}\,,
\end{eqnarray}
and the same condition as in eq.~(\ref{dVV=00003DdFF}) must be satisfied
for slow-roll inflation to be realised.

\section{Generalised Induced Gravity Inflation\label{sec:genIGr}}

In this section we analyse an induced gravity model of inflation \citep{Spokoiny:1984bd,Accetta:1985du,Futamase:1987ua}.
But we generalise the model by allowing for arbitrary powers of non-minimal
function and the spontaneous symmetry breaking potential. We are also
going to search for models which are consistent with the current observational
constraints of the scalar spectral index and tensor-to-scalar ratio.
The main purpose of this exercise is to make use of the equations
derived in the previous section.\footnote{Recently some works (see for example refs.~\citep{Kodama:2021yrm,Hyun:2022uzc})
have been published which analyse models that overlap with the ones
considered in this section in some parameter space. However, our main
goal is to derive analytical expressions of the constraints that can
be compared with observations.} 

The generalisation of the induced gravity model that we study can
be written as
\begin{eqnarray}
F & = & \xi\phi^{p}\,,\label{F-indG}\\
V & = & \frac{\lambda}{2q}\left(\phi^{2}-v^{2}\right)^{q}\,,\label{V-indG}
\end{eqnarray}
where $\left|p\right|$ and $q$ are of order 1. In addition $q$
must be an even, positive number in order for the potential to be
bounded from bellow.\footnote{Unbounded potential $V$ also results in an unbounded potential $U$
in eq.~(\ref{U}).} The $p=q=2$ case with $\phi\gg v$ has been studied extensively
in the literature (see. refs.~\citep{Spokoiny:1984bd,Accetta:1985du,Lucchin:1985ip,Futamase:1987ua,Pollock:1989vn,Kaiser:1993bq,Cervantes-Cota:1994qci,Fakir:1990iu,Salopek:1988qh,Kaiser:1994wj,Kaiser:1994vs}
for some early work on such models.). In this section our goal is
to investigate if inflation is possible in a broader range of parameters,
where $q$ is not necessarily equal to $p$ and the field $\phi$
rolls down the potential from both sides of the minimum. 

As $F$ must be positive to avoid instabilities, we fix $\xi>0$.
In that case it will be assumed that for \emph{odd} values of $p$
we are only interested in $\phi>0$ region.

In these class of models one further assumes that eventually the $\phi$
field settles down at the minimum of the potential $V$ with $\phi=v$.
At that point $F=1$ and we recover the action of General Relativity.
This requirement leads to the normalisation $\xi v^{p}=1$. We make
use of this relation to simplify the expressions by normalising the
$\phi$ field as
\begin{eqnarray}
x & \equiv & \frac{\phi}{v}\,.
\end{eqnarray}
The potential is symmetric under the transformation $\phi\rightarrow-\phi$.
Therefore, $x$ is positive.\footnote{If initially $\phi>0$, then it runs towards the $\phi\rightarrow v$
minimum and towards $\phi\rightarrow-v$ otherwise.} Using this definition eqs.~(\ref{F-indG}) and (\ref{V-indG}) can
be written as 
\begin{eqnarray}
F\left(x\right) & = & x^{p}\,,\\
V\left(x\right) & = & \frac{\lambda}{2q\xi^{2q/p}}\left(x^{2}-1\right)^{q}\,.
\end{eqnarray}
In the context of inflation the regime $x<1$ corresponds to a hilltop-like
inflation models and in the regime $x>1$ one is lead to the chaotic-like
models. We will use this terminology bellow to distinguish the two
regimes.

Remember, however, that in the Jordan frame the field does not necessarily
roll down the potential $V$, it can as well climb up the potential.
Hence, we need to impose additional conditions in order to guarantee
that the end point of the evolution is at $x=1$.

To that aim let us first write the slow-roll equation of motion (\ref{sr-eom-Jfrm})
for this model. In terms of $x$ it is given by  
\begin{eqnarray}
\frac{\dot{x}}{H} & \simeq & -\frac{4\xi^{2/p}x^{p-1}}{2+3p^{2}\xi^{2/p}x^{p-2}}\cdot\frac{\left(q-p\right)x^{2}+p}{x^{2}-1}\,.
\end{eqnarray}
As is discussed in the paragraph containing eq.~(\ref{Ugrad}), the
above slow-roll equation does not necessarily describe a field rolling
down towards the minimum of the potential $V\left(x\right)$. This
becomes evident if we take the initial conditions to be $x\ll1$,
which corresponds to the hilltop setup. In order for the field to
slow-roll towards $x=1$, the field velocity must be positive, $\dot{x}>0$.
 But this can be satisfied only in the case with $p>0$. Otherwise
the field climbs up the potential, away from the $x=1$ value and
towards $x=0$. In the opposite regime, where $x\gg1$, the field
rolls down the potential if the velocity is negative, $\dot{x}<0$.
Again, this leads to the condition $p>0$ if $q=p$, or $p<q$ if
$q\ne p$. Notice, that in the latter case it is consistent to consider
negative values of $p$. 

This behaviour becomes obvious if we write the Einstein frame scalar
field potential. According to eq.~(\ref{Ugrad}) 
\begin{eqnarray}
U_{,x} & = & 2U\left[\frac{\left(q-p\right)x^{2}+p}{x\left(x^{2}-1\right)}\right]\,.
\end{eqnarray}
We can readily notice that for very small $x$ the force is directed
towards the minimum of $U$ only for  $p>0$. While for large $x$
the same is true either for $p>0$, in the case of $p=q$, or for
$q>p$, in the case of $p\ne q$. These conditions are summarised
in the following equation
\begin{equation}
\begin{array}{cc}
p>0 & \text{ if }x\ll1\text{ (hilltop models)}\,,\\
p\le q & \text{ if }x\gg1\text{ (chaotic models)}\,.
\end{array}\label{pvals}
\end{equation}

Before separating the analysis into the hilltop and chaotic type regimes
let us first compute the general expression for the $\ev$ slow-roll
parameter. From eq.~(\ref{eu-gamma}) we find 
\begin{eqnarray}
\ev & \simeq & \frac{4p^{2}\xi^{2/p}x^{p-2}}{2+3p^{2}\xi^{2/p}x^{p-2}}\left[\frac{\left(\frac{q}{p}-1\right)x^{2}+1}{x^{2}-1}\right]^{2}\,.\label{eU-idGr}
\end{eqnarray}
As we know, the $\ev\ll1$ condition serves as a good proxy to indicate
the regime where $\eh 1\ll1$ (see eq.~(\ref{e1-sr})). The end of
inflation is then approximately given by the condition $\ev\simeq1$,
which we will use in the following computations.

\subsection{Hilltop Type Models}

\begin{figure}
\begin{centering}
\includegraphics[scale=0.4]{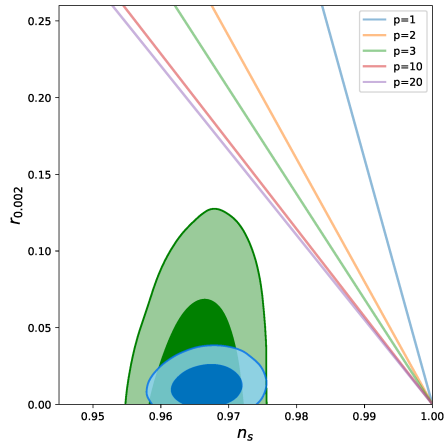}
\par\end{centering}
\caption{\label{fig:nsr-hilltop}The comparison of observational constraints
on scalar spectral index $n_{\mathrm{s}}$ and the tensor-to-scalar
ratio $r$ with predictions of the generalised induced gravity models
in the hilltop regime. The green contours represent $1\sigma$ and
$2\sigma$ constraints from ref.~\citep{Planck:2018jri} and the
blue contours add the new BICEP/Keck observations (see ref.~\citep{BICEP:2021xfz}
for details). $p>0$ is required for hilltop models. Note also, that
the presence of a line for some values of $r$ vs. $n_{\mathrm{s}}$
does not necessarily imply that those values can be actually realised
within a given model (which also holds for all the plots bellow).
To establish this one needs to compute the number of e-folds $\hat{N}_{*}$.
As the lines lie outside the observationally allowed region anyway,
we did not perform this computation.}
\end{figure}

In this regime we take the approximation
\begin{eqnarray}
x_{*} & \ll & 1\,,
\end{eqnarray}
where the asterisk denotes values when cosmological scales exit the
horizon. Within this approximation we can simplify the expression
for $\ev$ in eq.~(\ref{eU-idGr}) as
\begin{eqnarray}
\ev & \simeq & \frac{4p^{2}\xi^{2/p}x^{p-2}}{2+3p^{2}\xi^{2/p}x^{p-2}}\,.\label{eU-idGr-ht}
\end{eqnarray}
It depends on the value of $p$ if slow-roll inflation can be realised
on or not. Let us therefore consider first $p>2$. In this case $\ev$
monotonically decreases as we decrease $x$ and from eq.~(\ref{eU-idGr-ht})
it is clear that to satisfy the $\ev\ll1$ condition we need to impose
the bound
\begin{eqnarray}
p^{2}\xi^{2/p}x^{p-2} & \ll & 1\,.
\end{eqnarray}
Note, that from the definition of $F$ in eq.~(\ref{F-indG}), this
condition is equivalent to $F_{,\phi}^{2}\ll F$. Therefore, we can
use eq.~(\ref{hU-dF2-small}) to compute the second slow-roll parameter
$\hj$ 
\begin{eqnarray}
\hj & \simeq & -p\left(2-p\right)\xi^{2/p}x^{p-2}\,,
\end{eqnarray}
and approximate eq.~(\ref{eU-idGr-ht}) as
\begin{eqnarray}
\ev & \simeq & 2p^{2}\xi^{2/p}x^{p-2}\,.
\end{eqnarray}
Plugging the last two expressions into eq.~(\ref{ns-FV}) and (\ref{r-FV})
gives 
\begin{eqnarray}
n_{\mathrm{s}}-1 & \simeq & \frac{2-3p}{16p}r\,.\label{rns-hill}
\end{eqnarray}
As we can see in fig.~\ref{fig:nsr-hilltop} these models fall outside
the observationally allowed region.

The above relation is also applicable for $p=2$ and $p=1$ values
too. In the former case $\ev\simeq\mathrm{const}$ as long as $x\ll1$
is satisfied. Hence, slow-roll inflation, with $\ev\ll1$, is only
possible for 
\begin{eqnarray}
\xi & \ll & 1/2\,.\label{dF-p2}
\end{eqnarray}

In the case of $p=1$ the $p^{2}\xi^{2/p}x^{p-2}$ term is a decreasing
function of $x$. As we can clearly see from eq.~(\ref{eU-idGr-ht})
this implies 
\begin{eqnarray}
\ev & \overset{x\rightarrow0}{\rightarrow} & \frac{4}{3}\,,
\end{eqnarray}
which is a non-inflationary regime. Hence, the slow-roll inflation
can be realised only for field values 
\begin{eqnarray}
x & \gg & \frac{1}{2}\xi^{2}\,.\label{dF-p1}
\end{eqnarray}

Both expressions in eqs.~(\ref{dF-p2}) and (\ref{dF-p1}) imply
 $F_{,\phi}^{2}\ll F$. This justifies using an approximate relation
in eq.~(\ref{hU-dF2-small}) to compute $\hj$ and therefore eq.~(\ref{rns-hill})
to compute $n_{\mathrm{s}}\left(r\right)$ relation. Unfortunately,
all the hilltop type models lie outside the observationally allowed
region.

\subsection{Chaotic Type Models\label{subsec:Chaotic-Type}}

\begin{figure}
\begin{centering}
\subfloat[\label{fig:ch-p>q}Models with $p<q$ and $F_{,\phi}^{2}\ll F$.]{\begin{centering}
\includegraphics[scale=0.4]{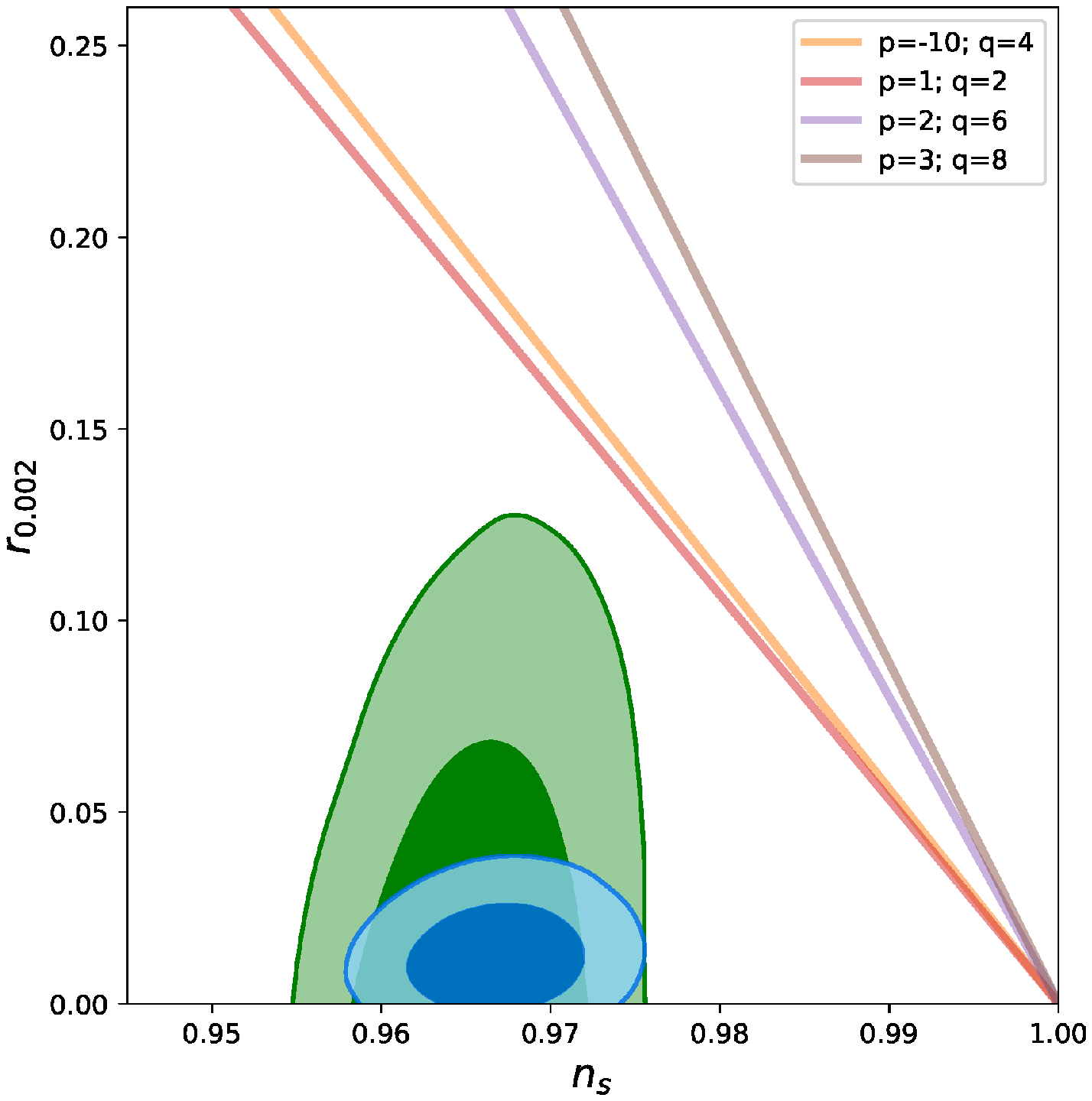}
\par\end{centering}
} ~ ~\subfloat[{\label{fig:nsr-chaotic-dFlarge}Models with $p=q$ and $F_{,\phi}^{2}\gg F$.
Thick sections of curves show $\hat{N}_{*}\in\left[50;60\right]$
range, where large points correspond to the $\hat{N}_{*}=60$ value.
The black dotted curves represent the results from exact numerical
simulations (see Appendix~\ref{sec:numerical}) in the same $\hat{N}_{*}$
range.}]{\begin{centering}
\includegraphics[scale=0.4]{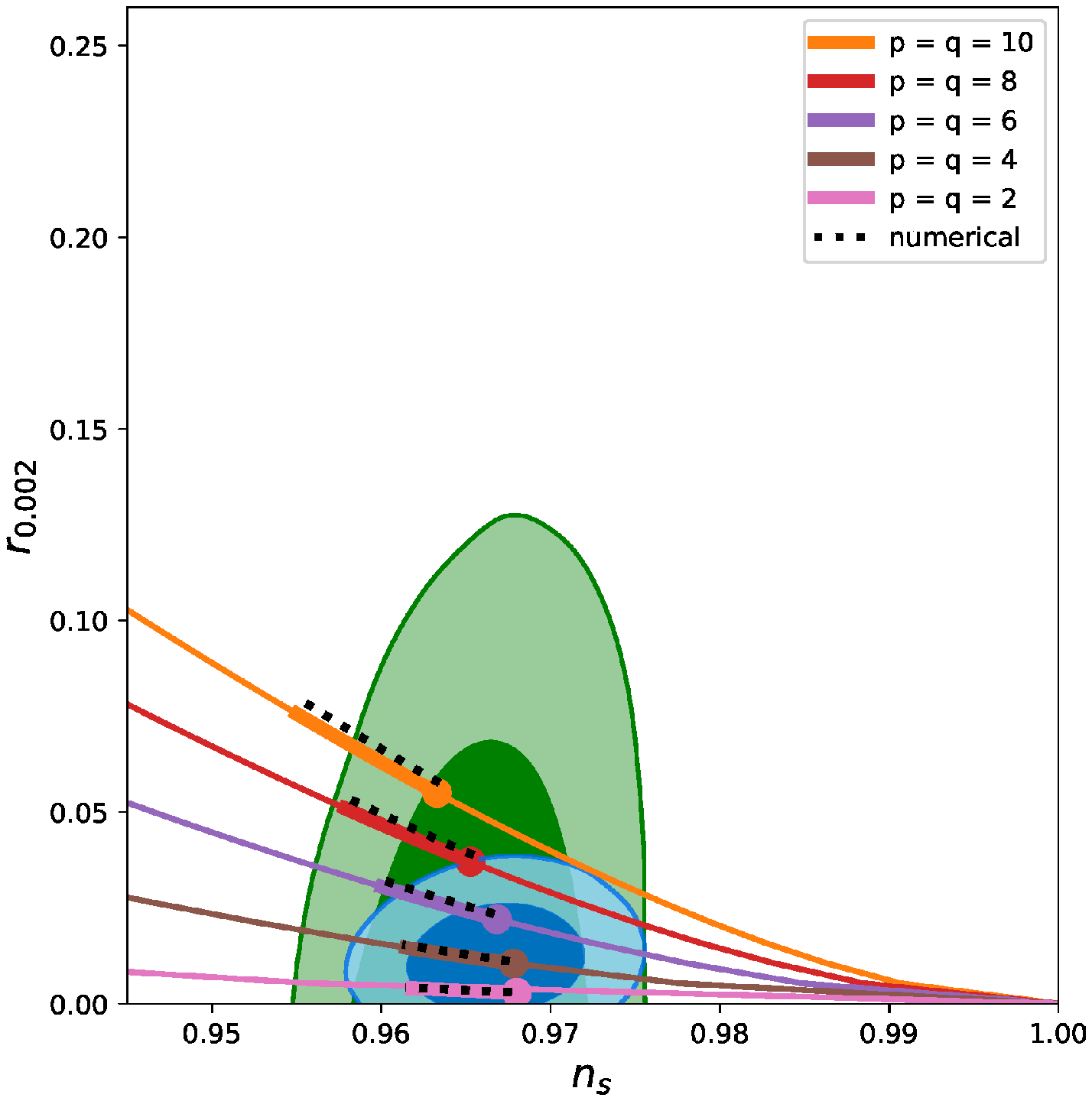}
\par\end{centering}
}
\par\end{centering}
\caption{Comparison of observational constraints of inflationary parameters
$n_{\mathrm{s}}$ and $r$ with chaotic type ($x_{*}\gg1$) models
of generalised induced gravity theories. The contours are the same
as in figure~\ref{fig:nsr-hilltop}.}
\end{figure}

We call ``chaotic'' models that satisfy
\begin{eqnarray}
x_{*} & \gg & 1\,.
\end{eqnarray}
The lowest order approximation of $\ev$ in eq.~(\ref{eU-idGr})
in terms of $x^{-2}$ depends on the relation between $p$ and $q$.
Let us consider first the case where $p\ne q$. As can be seen in
eq.~(\ref{pvals}) this case also permits negative values of $p$
and $\ev$ is approximately given by
\begin{eqnarray}
\ev & \simeq & \frac{4p^{2}\xi^{2/p}x^{p-2}}{2+3p^{2}\xi^{2/p}x^{p-2}}\left(\frac{q}{p}-1\right)^{2}\,.
\end{eqnarray}
In the limit $p^{2}\xi^{2/p}x^{p-2}\gg1$, which is equivalent to
$F_{,\phi}^{2}\gg F$, the first factor in the above expression is
$\sim4/3$. But for $p$ and $q$ of order 1, the second factor $\left(q/p-1\right)^{2}\sim\mathcal{O}\left(1\right)$.
Therefore slow-roll inflation can only be realised for models with
\begin{eqnarray}
p^{2}\xi^{2/p}x^{p-2} & \ll & 1\,.
\end{eqnarray}
As this condition is equivalent to  $F_{,\phi}^{2}\ll F$, we can
use an approximate expression of $\hj$ in eq.~(\ref{hU-dF2-small})
again 
\begin{eqnarray}
\hj & \simeq & \begin{cases}
qp^{2}\xi^{2/p}x^{p-4} & \text{for }p=2\\
\left(q-p\right)\left(2-p\right)\xi^{2/p}x^{p-2} & \text{for }p\ne2
\end{cases}\,,
\end{eqnarray}
to the lowest orders in $x^{-2}$ and $p^{2}\xi^{2/p}x^{p-2}$. 

We can readily notice that for $x\gg1$ the slow-roll parameter $\ev\gg\hj$.
Therefore using eqs.~(\ref{ns-FV}) and (\ref{r-FV}) we can write
a unified expression for all values of $p$ as
\begin{eqnarray}
n_{\mathrm{s}}-1 & \simeq & \frac{3\left(p-2\right)-2\left(q-2\right)}{16\left(q-p\right)}r\,.\label{rns-ch-p<q-dFsmall}
\end{eqnarray}
For $q=2$ the above equation reduces to $n_{\mathrm{s}}-1\simeq-3r/16$,
which is valid for any $p<q$. In fact this is the curve which gives
the smallest $r$ value for a given $n_{\mathrm{s}}$. Unfortunately,
as can be seen in fig.~\ref{fig:ch-p>q}, this parameter range is
already excluded by observations. 

\begin{figure}
\begin{centering}
\includegraphics[scale=0.5]{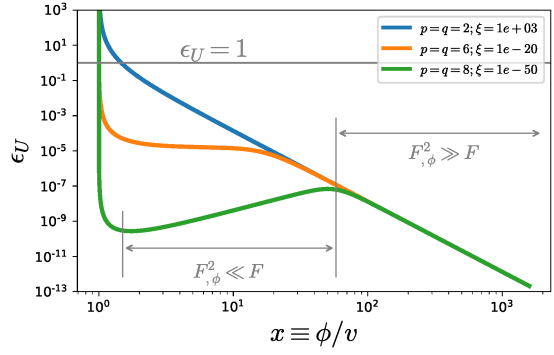}
\par\end{centering}
\caption{\label{fig:eU-ch_p=00003Dq}An illustration of the slow-roll parameter
$\protect\ev$ for chaotic type models with $p=q$. The regions that
correspond to $F_{,\phi}^{2}\ll F$ and $F_{,\phi}^{2}\gg F$ are
shown for the $p=8$ model. $\protect\ev$ can be approximated as
in eq.~(\ref{eU-ch-p=00003Dq-apx}) in those regions.}

\end{figure}

In the case of $p=q$ the $\ev$ function is shown in fig.~\ref{fig:eU-ch_p=00003Dq}
for several values of $p$ and $\xi$. To the lowest order in $x^{-2}$
we can approximate $\ev$ as 
\begin{eqnarray}
\ev & \simeq & \frac{4p^{2}\xi^{2/p}x^{p-2}}{2+3p^{2}\xi^{2/p}x^{p-2}}\cdot\frac{1}{x^{4}}\,.\label{eU-idGr-ch}
\end{eqnarray}
 It is then clear from the above result that inflation can be realised
in both regimes, for $F_{,\phi}^{2}\ll F$ (i.e. $p^{2}\xi^{2/p}x^{p-2}\ll1$)
and for $F_{,\phi}^{2}\gg F$. The expression for $\ev$ can be simplified
in both of these cases as
\begin{eqnarray}
\ev & \simeq & \begin{cases}
2p^{2}\xi^{2/p}x^{p-6} & \text{for }F_{,\phi}^{2}\ll F\\
\frac{4}{3}x^{-4} & \text{for }F_{,\phi}^{2}\gg F
\end{cases}\,.\label{eU-ch-p=00003Dq-apx}
\end{eqnarray}

Let us consider the case when cosmological scales exit the horizon
in the $F_{,\phi}^{2}\ll F$ regime first. Then we can approximate
the expression in eq.~(\ref{hU-dF2-small}) for large $x$ values
as 
\begin{eqnarray}
\hj & \simeq & \begin{cases}
\frac{6-p}{2p}x^{2}\ev & \text{for }p\ne6\\
\frac{1}{3}\ev & \text{for }p=6
\end{cases}\,,
\end{eqnarray}
where we also used eq.~(\ref{eU-ch-p=00003Dq-apx}). In the case
of $p=6$ and using eqs.~(\ref{ns-FV}) and (\ref{r-FV}) we find
\begin{eqnarray}
n_{\mathrm{s}}-1 & \simeq & -\frac{r}{6}\,.\label{ch-dFsmall-p=00003D6}
\end{eqnarray}
These values also lie outside the observationally allowed region.
In the case of $p\ne6$, $\hj\gg\ev$ and eq.~(\ref{ns-FV}) is
approximated by
\begin{eqnarray}
n_{\mathrm{s}}-1 & \simeq & \frac{p-6}{p}x^{2}\ev\,.\label{ch-dFsmall-p=00003Dq}
\end{eqnarray}
The $1-n_{s}\simeq10^{-2}$ constraint can only be satisfied for $x\sim1$.
But for these large values the $x\gg1$ condition, that was used to
derive eq.~(\ref{eU-idGr-ch}), is violated. That is, the equations
above, where the approximation $F_{,\phi}^{2}\ll F$ was used, are
applicable only for much smaller values of $1-n_{\mathrm{s}}$ than
what is allowed by observations.

Therefore, we are only left with the region, in which the $F_{,\phi}^{2}\gg F$
(i.e. $p^{2}\xi^{2/p}x^{p-2}\gg1$) approximation holds. In this limit
$\ev$ in eq.~(\ref{eU-idGr-ch}) is given by
\begin{eqnarray}
\ev & \simeq & \frac{4}{3x^{4}}\,,\label{eU-ch-dFlarge}
\end{eqnarray}
which is independent of $p$. The second slow-roll parameter $\hj$,
on the other hand, can be approximated by eq.~(\ref{hU-dF2_large}).
For the current model and using $x\gg1$, this equation can be written
as 
\begin{eqnarray}
\hj & \simeq & \frac{4}{p}\sqrt{\frac{\ev}{3}}+\frac{4}{p}\ev\,,\label{hFV-ch-dFlarge}
\end{eqnarray}
which leads to 
\begin{eqnarray}
n_{\mathrm{s}}-1 & \simeq & -\frac{2}{p}\sqrt{\frac{r}{3}}-\left(\frac{4}{p}+1\right)\frac{r}{8}\,.\label{rns-ch-dFlarge}
\end{eqnarray}

As we can see in fig.~\ref{fig:nsr-chaotic-dFlarge} these models
cross the BICEP/Keck region with $q=p$ values up to $10$. However,
we need to determine if this happens at the right number of e-folds
before the end of inflation. To that end we can write eq.~(\ref{N-SR-Jfrm-2})
in terms of the rescaled variable $x$ as 
\begin{eqnarray}
N & \simeq & \frac{1}{8\xi^{2/p}p}\intop_{x_{\mathrm{end}}}^{x_{*}}\left(2+3p^{2}\xi^{2/p}x^{p-2}\right)\left(x^{2}-1\right)\frac{\mathrm{d}x^{2}}{x^{p}}\,,\label{N-indGr-ch-q=00003Dp}
\end{eqnarray}
which can be readily integrated, leading to 
\begin{eqnarray}
N_{*} & \simeq & \begin{cases}
\frac{3}{4}\left(\frac{1}{6\xi}+1\right)\left(x_{*}^{2}-x_{\e}^{2}+\ln\frac{x_{\e}^{2}}{x_{*}^{2}}\right)\,, & \text{for }p=2\\
\frac{1}{16\xi^{2}}\left(x_{*}^{-2}-x_{\e}^{-2}-\ln\frac{x_{\e}^{2}}{x_{*}^{2}}\right)+\frac{3}{2}\left(x_{*}^{2}-x_{\e}^{2}+\ln\frac{x_{\e}^{2}}{x_{*}^{2}}\right)\,, & \text{for }p=4\\
\frac{1}{2p\xi^{2/p}}\left(\frac{x_{*}^{4-p}-x_{\e}^{4-p}}{4-p}-\frac{x_{*}^{2-p}-x_{\e}^{2-p}}{2-p}\right)+\frac{3p}{8}\left(x_{*}^{2}-x_{\mathrm{\e}}^{2}+\ln\frac{x_{\e}^{2}}{x_{*}^{2}}\right)\,, & \text{otherwise}
\end{cases}\,.
\end{eqnarray}
We have already assumed that when cosmological scales exit the horizon
the $F_{,\phi}^{2}\gg F$ condition holds. In the case $p=2$, this
is equivalent to saying that $4\xi\gg1$. For other values of $p$
this condition implies $p^{2}\xi^{2/p}x_{*}^{p-2}\gg1$. Thus, we
can safely neglect the $\left(p^{2}\xi^{2/p}x_{*}^{p-2}\right)^{-1}$
terms, as compared to $x_{*}^{2}\gg1$ term or 1, from the above expression.
This, however, does not imply that we can neglect $\left(p^{2}\xi^{2/p}x_{\e}^{p-2}\right)^{-1}$
terms. In the models with $p>2$, $\xi$ is allowed to be small and
still satisfy $p^{2}\xi^{2/p}x_{*}^{p-2}\gg1$ condition, in principle.

However, as we will see shortly, in practice observational constraints
push $\xi$ to be large and we can neglect $\left(p^{2}\xi^{2/p}x_{\e}^{p-2}\right)^{-1}$
terms too. Making use of $x_{*}\gg x_{\e}$ we can eventually approximate
$N_{*}$ by 
\begin{eqnarray}
N_{*} & \simeq & \frac{3p}{8}\left(x_{*}^{2}+\ln\frac{x_{\e}^{2}}{x_{*}^{2}}\right)\,.
\end{eqnarray}
The logarithmic term is of the same order as the e-fold shift number
in eq.~(\ref{thN-SR-Jfrm}). Plugging that equation into the above
we find the e-fold number in the Einstein frame 
\begin{eqnarray}
\hat{N}_{*} & \simeq & \frac{p}{8}\left(3x_{*}^{2}+5\ln\frac{x_{\e}^{2}}{x_{*}^{2}}\right)\,.\label{hatN-ch-p=00003Dq}
\end{eqnarray}
The logarithmic term can introduce a sizeable correction to $\hat{N}_{*}$
and therefore cannot be neglected. Unfortunately, this means that
we need to solve a transcendental equation for $x_{*}$, which we
do numerically.

But before solving for $x_{*}$ we need to find $x_{\e}$. With the
assumption $p^{2}\xi^{2/p}\gg1$ (or $\xi\gg p^{-p}$) it can be easily
computed from eq.~(\ref{eU-idGr}). Plugging in $q=p$ into that
expression and taking $\left.\ev\right|_{\e}=1$ we find
\begin{eqnarray}
x_{\e} & \simeq & \sqrt{1+\frac{2}{\sqrt{3}}}\simeq1.35\,.
\end{eqnarray}

Now we can find $x_{*}$ from eq.~(\ref{hatN-ch-p=00003Dq}) corresponding
to $\hat{N}_{*}=50$ and $60$ e-folds of inflation. Plugging that
value into eqs.~(\ref{eU-ch-dFlarge}), (\ref{hFV-ch-dFlarge}) and
(\ref{rns-ch-dFlarge}) we are able to compare the results with observations.
They are shown in figure~\ref{fig:nsr-chaotic-dFlarge}. As we can
see practically only $p=q=2$, $4$ and $6$ values are allowed.

\todo[inline]{Find the values of $\lambda$ from $A_{s}$ constraints that fit the
data.}

\subsection{Slow-Roll Inflation for $x\simeq1$}

\begin{figure}
\begin{centering}
\includegraphics[scale=0.4]{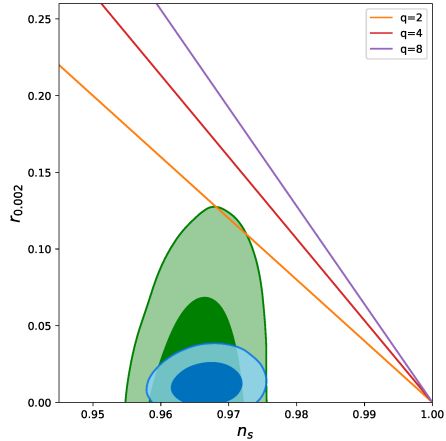}
\par\end{centering}
\caption{\label{fig:nsr-x1}Comparison of observational constraints with $x_{*}\simeq1$
models. The contours are the same as in figure~\ref{fig:nsr-hilltop}.}
\end{figure}

So far we have considered models for which cosmological scales exit
the horizon either when $x_{*}\ll1$ or $x_{*}\gg1$. In this section
we look for observationally acceptable models with $x_{*}\simeq1$.
To that aim, let us define
\begin{eqnarray}
\delta & \equiv & 1-x^{2}\,,
\end{eqnarray}
where $\left|\delta\right|\ll1$.

Before doing the analysis note that in this setup the constraints
on $p$ in eq.~(\ref{pvals}) do not apply. Indeed, close to $x\simeq1$
the gravitational force is always directed towards the minimum of
the potential, as can be confirmed by writing
\begin{eqnarray}
U_{,x} & = & 2U\left[-\frac{q-\left(q-p\right)\delta}{\left(1-\frac{1}{2}\delta\right)\delta}\right]\,.
\end{eqnarray}
For $x\lesssim1$ $\delta$ is positive, which makes $U_{,x}$ negative
and vice versa. 

Taking the first two terms in the expansion of $\ev$ in terms of
$\delta^{-1}$ we can write
\begin{eqnarray}
\ev & \simeq & \frac{4p^{2}\xi^{2/p}}{2+3p^{2}\xi^{2/p}}\left(\frac{q}{p}\right)^{2}\frac{1}{\delta^{2}}\,.
\end{eqnarray}
It follows that slow-roll inflation, with $\ev\ll1$, can be realised
only for
\begin{eqnarray}
p^{2}\xi^{2/p} & \ll & 1\,,
\end{eqnarray}
that is $F_{,\phi}^{2}\ll F$. Using this fact the above equation
for $\ev$ can be simplified:  
\begin{eqnarray}
\ev & \simeq & \frac{2q^{2}\xi^{2/p}}{\delta^{2}}\,.
\end{eqnarray}
Applying the same approximations to the expression of $\hj$ in eq.~(\ref{hj-def})
we get 
\begin{eqnarray}
\hj & \simeq & \frac{4q\xi^{2/p}}{\delta^{2}}\,.
\end{eqnarray}
Plugging in the last two results into eqs.~(\ref{ns-FV}) and (\ref{r-FV})
one finally arrives at
\begin{eqnarray}
n_{\mathrm{s}}-1 & \simeq & -\frac{q+2}{8q}r\,.\label{rns-x=00003D1}
\end{eqnarray}
The smallest $r$ value for a given $n_{\mathrm{s}}$ is achieved
with $q=2$. However, as can be seen in fig.~\ref{fig:nsr-x1}, even
for these values the predictions lie outside the $2\sigma$ contour
of the BICEP/Keck results. 

Our results of section \ref{sec:genIGr} are summarised in table~\ref{tab:regimes}.

\begin{table}

\begin{centering}
\begin{tabular}{|l|c|l|l|}
\cline{3-4} \cline{4-4} 
\multicolumn{1}{l}{\multirow{1}{*}{}} & \multirow{1}{*}{} & $F_{,\phi*}^{2}\ll F_{*}$ & $F_{,\phi*}^{2}\gg F_{*}$\tabularnewline
\hline 
\multicolumn{2}{|l|}{Hilltop ($x_{*}\ll1$)} & eq.~(\ref{rns-hill}): $n_{\mathrm{s}}-1\simeq\frac{2-3p}{16p}r$  & no slow-roll\tabularnewline
\hline 
\multirow{3}{*}{Chaotic ($x_{*}\gg1$)} & $q>p$ & eq.~(\ref{rns-ch-p<q-dFsmall}): $n_{\mathrm{s}}-1\simeq\frac{3\left(p-2\right)-2\left(q-2\right)}{16\left(q-p\right)}r$  & no slow-roll\tabularnewline
\cline{2-4} \cline{3-4} \cline{4-4} 
 & $q=p=6$ & eq.~(\ref{ch-dFsmall-p=00003D6}): $n_{\mathrm{s}}-1\simeq-\frac{r}{6}$  & \multirow{2}{*}{eq.~(\ref{rns-ch-dFlarge}): $n_{\mathrm{s}}-1\simeq-\frac{2}{p}\sqrt{\frac{r}{3}}-\left(\frac{4}{p}+1\right)\frac{r}{8}$ }\tabularnewline
\cline{2-3} \cline{3-3} 
 & $q=p\ne6$ & eq.~(\ref{ch-dFsmall-p=00003Dq}): $n_{\mathrm{s}}-1\simeq\frac{p-6}{p}x_{*}^{2}\ev$ & \tabularnewline
\hline 
\multicolumn{2}{|l|}{$x_{*}\simeq1$} & eq.~(\ref{rns-x=00003D1}): $n_{\mathrm{s}}-1\simeq-\frac{q+2}{8q}r$  & no slow-roll\tabularnewline
\hline 
\end{tabular}
\par\end{centering}
\caption{\label{tab:regimes}The summary of observational predictions of the
generalised induced gravity inflation models in different regimes.
In this table ``no slow-roll'' signifies the absence of $\protect\ev\ll1$
region. The only models that are compatible with observations are
chaotic type models with $p=q\protect\leq6$ in the regime $F_{,\phi*}^{2}\gg F_{*}$
(see fig.~\ref{fig:nsr-chaotic-dFlarge}).}

\end{table}

\section{Summary and Conclusions}

In this work we consider inflation models within the framework of
scalar-tensor theories of gravity. The most straightforward way to
analyse such models and confront them with observations is to transform
the action into the Einstein frame. But this is not always desirable.
For example, after the transformation the matter sector directly couples
to the scalar field, even if no such coupling is introduced in the
Jordan frame. For certain applications this might actually complicate
the analysis. In those cases, it would be convenient to have a formalism
where inflation observables can be calculated only relying on Jordan
frame quantities. In this work we derive the requirements that Jordan
frame quantities must satisfy to provide slow-roll inflation. We also
derive Jordan frame flow and slow-roll parameters which must be small,
and write inflation observables in terms of those parameters.

The central idea of our method is to utilise the fact that conformal
transformation corresponds to the change of units of measure \citep{Dicke:1961gz}.
We assume to slice the spacetime into spacelike hypersurfaces and
keep that slicing fixed. It allows us to map the relations defined
in the Einstein frame to the Jordan frame on the same slice by changing
the units of measure (performing the conformal transformation). In
particular, we take the conditions required for inflation and more
stringent conditions for slow-roll inflation, which are most clearly
defined in the Einstein frame, and map them to the Jordan frame.

We also make use of another convenient fact that the scalar and tensor
metric perturbation spectra in single field slow-roll inflation models
can be conveniently written in terms of homogeneous quantities only.
Hence, only the homogeneous relations need to be mapped from the Einstein
to the Jordan frame in order to derive observable model predictions.

Using this method we first derive the conditions for inflation. It
has been noted in various places in the literature (see e.g. \citep{Faraoni:2004pi})
that those conditions in the Jordan frame might look somewhat counterintuitive
from the Einstein frame perspective. For example, the Jordan frame
scale factor does not have to be accelerating, as was demonstrated
in eq.~(\ref{a-infl}). Or the Hubble flow parameter $\ej 1$ does
not have to be smaller than 1 (see eq.~(\ref{inf-cond-Jfrm})). This
has to be kept in mind when doing computations that are related to
the end of inflation.

However, once we impose a much stronger requirement, that the Einstein
frame Hubble-flow parameter $\eh 1\ll1$ (see eq.~(\ref{e1-def})
for the definition), the conditions become more stringent. In that
case some of the familiar expressions in the Einstein frame can be
carried over to the Jordan one. In particular eqs.~(\ref{sr1-Jrm-v2-t1})
and (\ref{dV-sr}), which for convenience we rewrite here, are
\begin{align*}
\frac{1}{2}\dot{\phi}^{2} & \ll V\,,\\
\frac{\dot{V}}{HV} & \ll1\,.
\end{align*}

In the Jordan frame we have an additional function $F$. Which enlarges
the number of conditions. To find a compact parametrisation for them
we introduced a number of ``flow parameters'', in analogy to the
``Hubble-flow parameters''. These are $\ej i$, $\theta_{i}$ and
$\gamma^{2}$ defined in eqs.~(\ref{e1-Jfrm}) - (\ref{gm-def}).
We demonstrated in this work that the requirement $\eh i\ll1$ translates
into the Jordan frame as
\[
\ej 1,\:\gamma^{2},\:\left|\gamma_{2}\right|,\:\left|\theta_{1}\right|,\:\left|\theta_{1}\theta_{2}\right|\ll1\,.
\]
The above realtions are derived in eqs.~(\ref{e1-ll-1}), (\ref{e1-EJfrms-v2}),
(\ref{g2-ll-1}), (\ref{th-small}) and (\ref{th2-constr}) respectively. 

Only two of the three sets of parameters $\ej i$, $\theta_{i}$ and
$\gamma_{i}$ are needed to describe the system. And we found that
using $\theta_{i}$ and $\gamma_{i}$ results in the most compact
equations. However, when computing inflation observables we provide
two combinations. In eqs.~(\ref{As-Jf})-(\ref{r-Jf}) the scalar
spectral index, scalar spectral tilt and tensor-to-scalar ratio, $A_{\mathrm{s}}$,
$n_{\mathrm{s}}$ and $r$ respectively, are written as functions
of $\gamma_{i}$ and $\theta_{i}$
\begin{align*}
A_{\mathrm{s}} & \simeq\frac{V}{24\pi^{2}F^{2}\gamma^{2}}\,,\\
n_{\mathrm{s}}-1 & \simeq-2\left(\gamma^{2}+\gamma_{2}-\theta_{1}\theta_{2}\right)\,,\\
r & \simeq16\gamma^{2}
\,.\end{align*}
 And in eqs.~(\ref{As-ev})-(\ref{r-ev}), they are written as functions
of $\ej i$ and $\theta_{i}$.

We also derive the slow-roll equations in the Jordan frame. As described
above, we take the Einstein frame version of slow-roll conditions
in eqs.~(\ref{sr1-Ef-t}) and (\ref{sr2-Ef-t}) and after the conformal
transformation they can be written as in eqs.~(\ref{sr1-Jrm-v1})
and (\ref{sr2-Jrm-v1}). Taking into account all the smallness conditions
discussed above we finally arrive at eqs.~(\ref{sr-H-Jfrm}) and
(\ref{sr-eom-Jfrm})
\begin{align*}
H^{2} & \simeq\frac{V}{3F}\,,\\
3H\dot{\phi} & \simeq\frac{2VF_{,\phi}-FV_{,\phi}}{F^{2}\mathcal{K}}\,,
\end{align*}
which are the Jordan frame versions of the slow-roll Friedman equation
and equation of motion. These approximate relations can be applied
as long as slow-roll conditions $\ev,\,\left|\hj\right|\ll1$ are
satisfied, where the slow-roll parameters are defined in eqs.~(\ref{eu-gamma})
and (\ref{hj-def})
\begin{align*}
\ev & =\frac{1}{2\mathcal{K}}\left(\frac{V_{,\phi}}{V}-2\frac{F_{,\phi}}{F}\right)^{2}\,,\\
\hj & =\frac{1}{\mathcal{K}}\left[2\frac{F_{,\phi\phi}}{F}-\frac{V_{,\phi\phi}}{V}-2\frac{F_{,\phi}^{2}}{F^{2}}+\frac{V_{,\phi}^{2}}{V^{2}}+\frac{\mathcal{K}_{,\phi}}{2\mathcal{K}}\left(\frac{V_{,\phi}}{V}-2\frac{F_{,\phi}}{F}\right)\right]\,,
\end{align*}
where $\ev$ is the same slow-roll parameter as in the Einstein frame
but expressed in terms of the Jordan frame quantities. The dependence
of inflation observables on $\ev$ and $\hj$ is given in eqs.~(\ref{As-FV})-(\ref{r-FV})
\begin{eqnarray*}
A_{\mathrm{s}} & \simeq & \frac{V}{24\pi^{2}F^{2}\ev}\,,\\
n_{\mathrm{s}}-1 & \simeq & -2\left(\ev+\hj\right)\,,\\
r & \simeq & 16\ev\,.
\end{eqnarray*}

When computing numerical values of the above parameters, another aspect
that has to be taken into account is the difference between the number
of e-folds of inflation as defined with respect to $\hat{a}$ and
$a$, i.e. the scale factors in the Einstein and Jordan frames respectively.
This difference is proportional to the logarithm of $F$, as shown
in eq.~(\ref{thN-SR-Jfrm})
\[
\hat{N}=N+\frac{1}{2}\ln\frac{F_{\e}}{F}\,.
\]
Although the e-fold shift number $\ln\sqrt{F_{\e}/F}$ is logarithmic,
it can be substantial and therefore generically cannot be neglected,
as we demonstrate this in the case of induced gravity models of inflation.
When comparing with observations, we used $\hat{N}$ from 50 to 60,
which are the values used by the Planck team.

Finally, we should also mention another known fact, that in the Jordan
frame the field does not necessarily roll down the potential $V\left(\phi\right)$
even in slow-roll. As the strength and the direction of the gravitational
force depends on $F\left(\phi\right)$, the field $\phi$ can as well
climb up that potential. This can be readily understood by looking
at the gradient of the scalar field potential in the Einstein frame
in eq.~(\ref{Ugrad}). The direction of the gravitational force is
no longer determined solely by the gradient of $V\left(\phi\right)$,
but in combination with the gradient of $F\left(\phi\right)$. This
has to be taken into account when considering models of inflation,
as we also demonstrate in our example model in section~\ref{sec:genIGr}.

In that section we consider a generalised induced gravity model. The
main idea of induced gravity theories is to generate gravity by a
spontaneous symmetry breaking \citep{Sakharov:1967pk,OHanlon:1972xqa,Zee:1978wi,Smolin:1979uz}.
That is, General Relativity is recovered after the $\phi$ field settles
at its VEV, which is determined by the potential $V\left(\phi\right)$.
To make sure that the field slowly rolls towards the minimum of $V\left(\phi\right)$
one needs to constraint possible functional forms of $F\left(\phi\right)$.

The main goal of section~\ref{sec:genIGr} is to apply some of our
results to a concrete model. Making use of Jordan frame quantities
only we analyse the model specified in eqs.~(\ref{F-indG}) and (\ref{V-indG})
and look for observationally acceptable parameter space. We found
that, among the many regions where slow roll inflation could be realised,
only the case with $p=q=2,\,4$ and $6$ in the ``chaotic type''
regime (where $\left|\phi_{*}\right|\gg\left|v\right|$) fall within
$2\sigma$ region of the newest BICEP/Keck results \citep{BICEP:2021xfz}
(see figure~\ref{fig:nsr-chaotic-dFlarge}).

Finally, to validate our method we performed numerical simulations
in which we solved exact perturbation equations in the Einstein frame
and compared them with our Jordan frame slow-roll approximations.
As can be seen in the Appendix~\ref{sec:numerical} and figure~\ref{fig:nsr-chaotic-dFlarge}
the agreement is very good indeed.

In this work we neglected a possible contribution from matter fields
to the dynamics of the system. We also analysed single field models
only. These simplifications allowed a simple check of the above results
by transforming the action into the Einstein frame and performing
exact computations numerically. In the future we plan to extend our
formalism by including those complications, such as matter fields
and considering multi-field models, which will bring us closer to
the real motivation of this work.

\appendix

\section{Comparison of Flow Parameters with Numerical Simulations\label{sec:numerical}}

\begin{figure}
\subfloat[\label{fig:e1}Numerical check of eq.~(\ref{e1-gamma}), where $\gamma$
is approximated by eq.~(\ref{gamma2-theta-apx}).]{\begin{centering}
\includegraphics[scale=0.3]{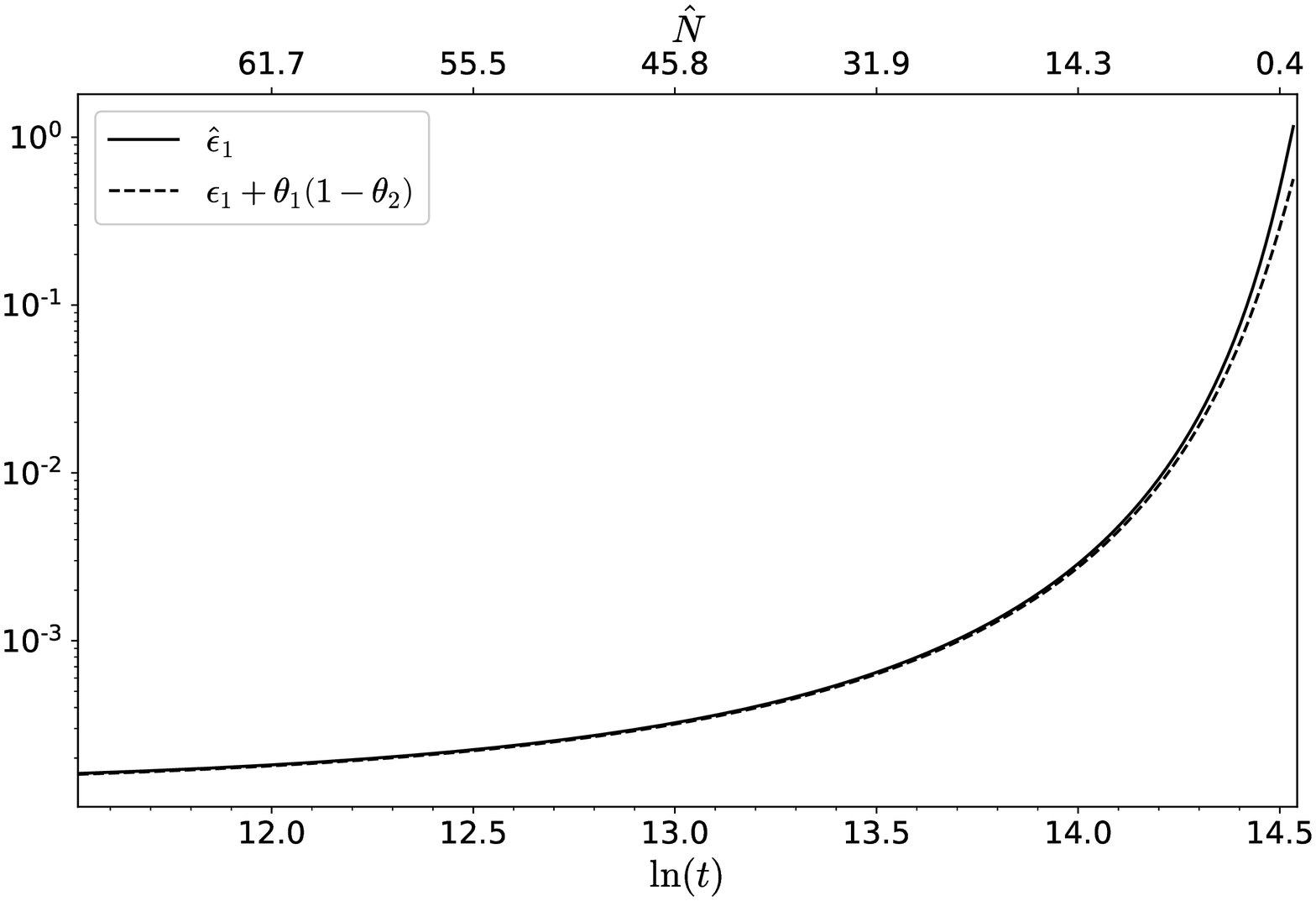}
\par\end{centering}
}~\subfloat[\label{fig:e2}Numerical check of eq.~(\ref{e2-EJfrms}).]{\begin{centering}
\includegraphics[scale=0.3]{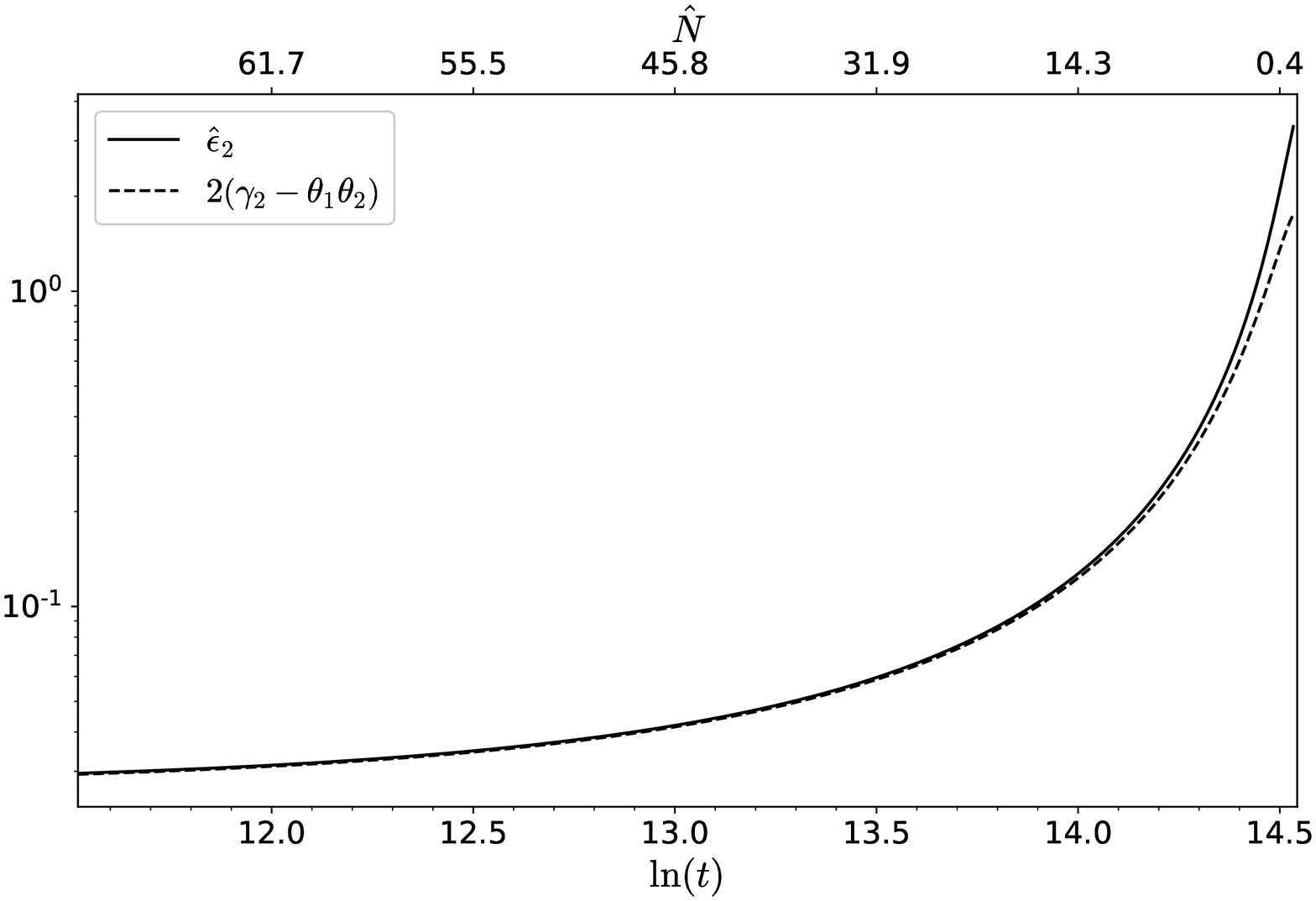}
\par\end{centering}
}

\subfloat[\label{fig:As}Numerical check of eq.~(\ref{As-Jf}).]{\begin{centering}
\includegraphics[scale=0.3]{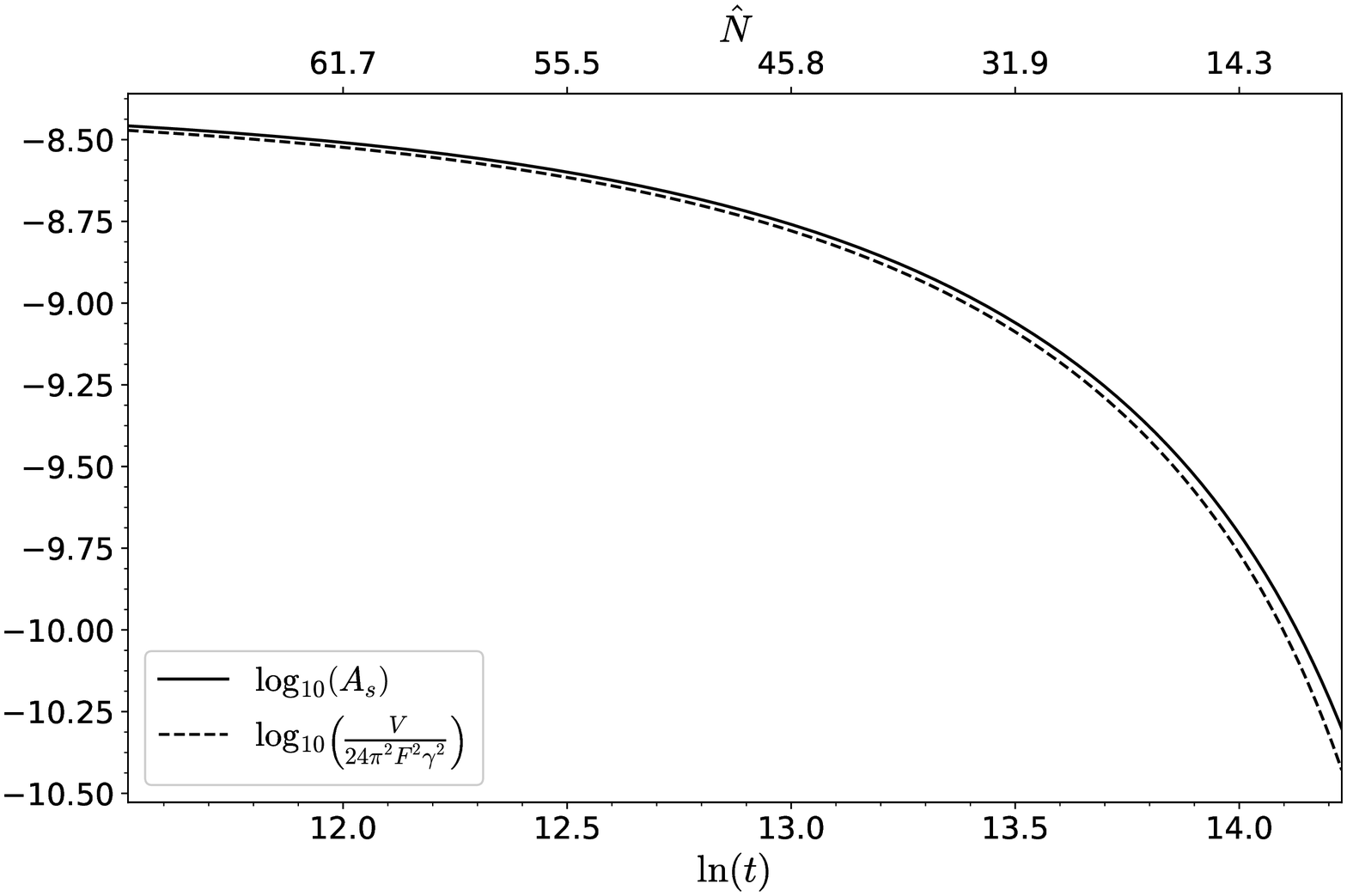}
\par\end{centering}
}~\subfloat[\label{fig:ns}Numerical check of eq.~(\ref{ns-Jf}).]{\centering \includegraphics[scale=0.3]{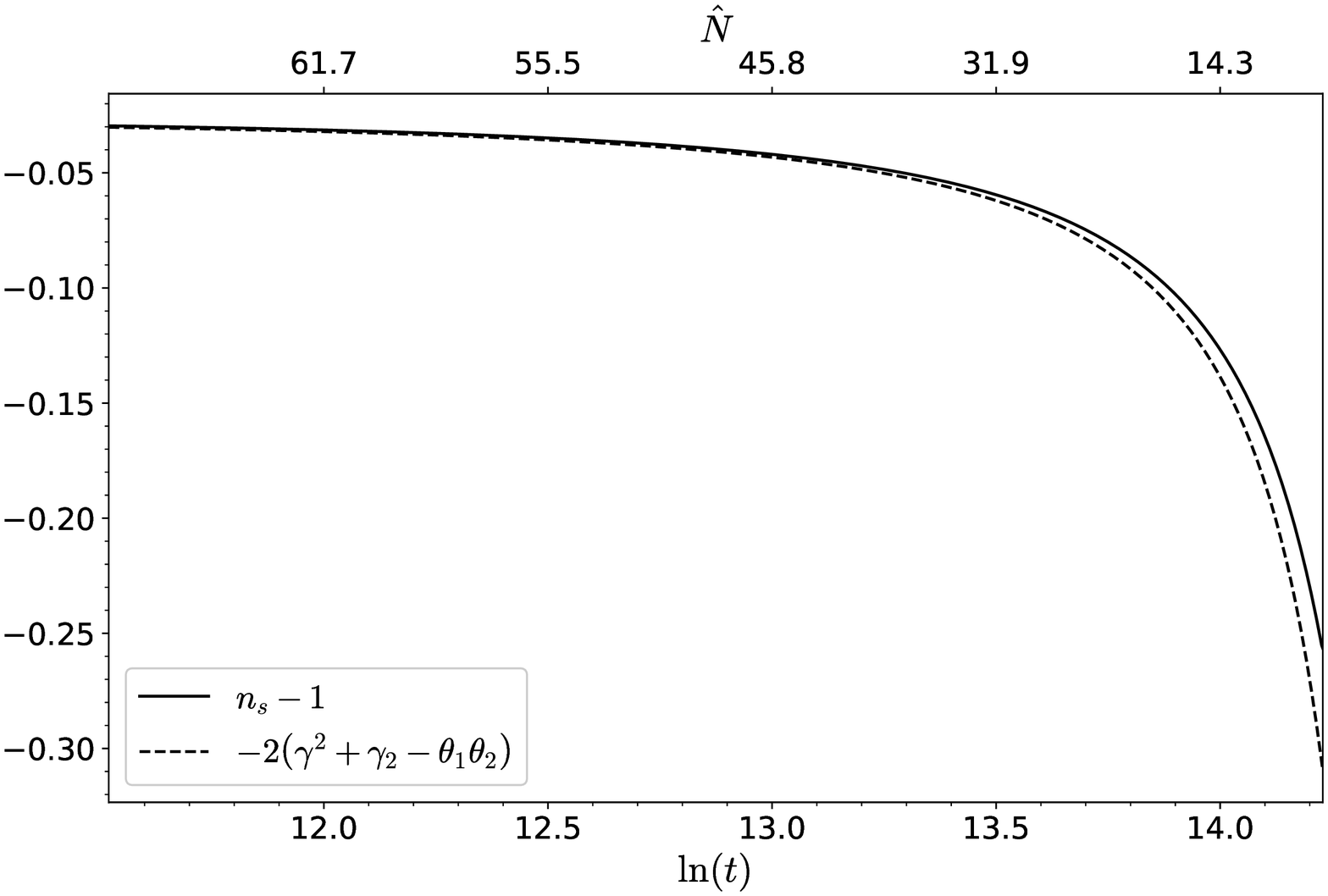}

}

~ \subfloat[\label{fig:r}Numerical check of eq.~(\ref{r-Jf}).]{\begin{centering}
\includegraphics[scale=0.3]{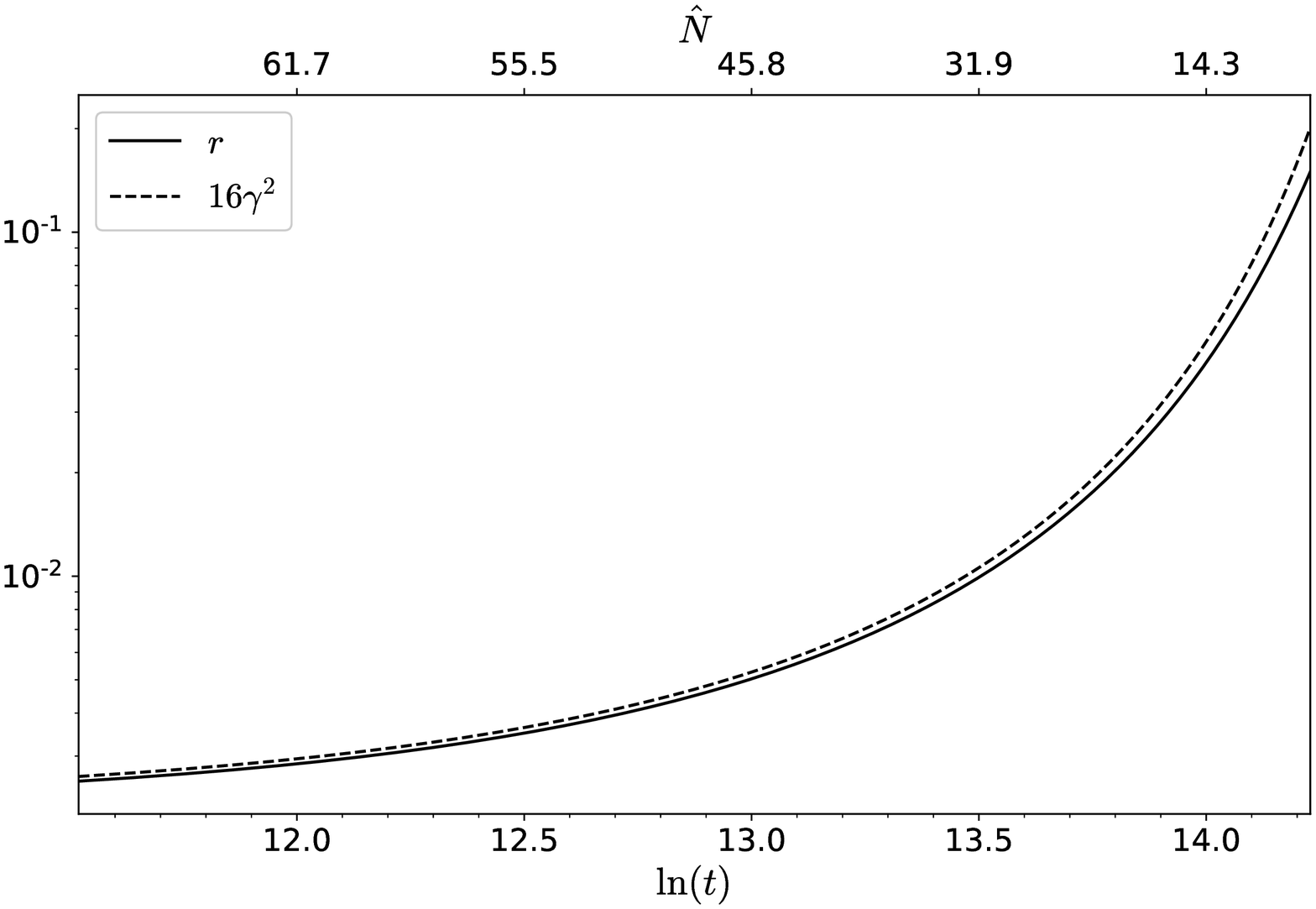}
\par\end{centering}
}\caption{\label{fig:comparison}The comparison of exact, numerical Einstein
frame results and approximate relations derived in this work. We used
the model defined in eqs.~(\ref{F-indG}) and (\ref{V-indG}) with
$p=q=2$. These functions are plotted against the time $t$, which
enumerates spatial slices and are shown on the lower horizontal axis.
The upper horizontal axis shows the corresponding e-fold number $\hat{N}\left(t\right)$
in the Einstein frame for the reference.}
\end{figure}

To see how well our approximations perform we compared them with exact
numerical solutions. To that purpose we used the induced gravity model
in eqs.~(\ref{F-indG}) and (\ref{V-indG}) with $p=q=2$ and $\phi_{*}/v\gg1$.
As shown in figure~\ref{fig:nsr-chaotic-dFlarge} this model conforms
to observations very well.

The numerical simulations are performed in both frames. In the case
of Einstein frame, we first transform the Jordan frame action in eq.~(\ref{S-Jfr})
using the conformal transformation in eq.~(\ref{conf-tr}). This
results in a scalar field with a non-canonical kinetic function as
in eq.~(\ref{S-Ef}) and background equations (\ref{eom-tau}), (\ref{Fdm-eq})
and (\ref{cont-eq}). Using eq.~(\ref{N-def-Ef}) they can be also
expressed in terms of the number of e-folds $\hat{N}$ 
\begin{align}
 & \frac{\mathrm{d}^{2}\phi}{\mathrm{d}\hat{N}^{2}}-\left(3-\frac{1}{\hat{H}}\frac{\mathrm{d}\hat{H}}{\mathrm{d}\hat{N}}\right)\frac{\mathrm{d}\phi}{\mathrm{d}\hat{N}}+\frac{1}{2}\frac{\mathcal{K}_{,\phi}}{\mathcal{K}}\left(\frac{\mathrm{d}\phi}{\mathrm{d}\hat{N}}\right)^{2}+\frac{U_{,\phi}}{\mathcal{K}\hat{H}^{2}}=0\,,\label{num-EoM}\\
 & \hat{H}^{2}=\frac{U\left(\phi\right)}{3-\frac{1}{2}\mathcal{K}\left(\frac{\mathrm{d}\phi}{\mathrm{d}\hat{N}}\right)^{2}}\,,\label{num-H2}\\
 & \frac{\mathrm{d}\hat{H}}{\mathrm{d}\hat{N}}=\frac{1}{2}\mathcal{K}\hat{H}\left(\frac{\mathrm{d}\phi}{\mathrm{d}\hat{N}}\right)^{2}\,,\label{num-dH}
\end{align}
where the kinetic term is given by 
\begin{equation}
\mathcal{K}=\left(\frac{1}{\xi}+6\right)\frac{1}{\phi^{2}}\,,
\end{equation}
and the lapse function $\hat{\n}=\sqrt{F}=\sqrt{\xi}\phi$.

Alongside homogeneous equations we also integrate equations for perturbations.
The curvature perturbation $\hat{\mathcal{R}}_{k}$ is related to
the field perturbations $\delta\phi_{k}$ by \citep{Hwang:1996xh}
\begin{equation}
\hat{\mathcal{R}}_{k}=-\frac{\hat{u}_{k}}{\hat{a}\frac{\textrm{d}\phi}{\textrm{d}\hat{N}}\sqrt{\mathcal{K}}}\,,
\end{equation}
where $\hat{u}_{k}\equiv\hat{a}\sqrt{\mathcal{K}}\delta\phi_{k}$
is the Mukhanov-Sasaki variable. The evolution of $\hat{\mathcal{R}}_{k}$
is governed by the equation 
\begin{equation}
\ddot{\hat{\mathcal{R}}}_{k}+\left[\left(3+\hat{\epsilon}_{2}\right)\sqrt{F}-\frac{1}{2}\frac{\dot{F}}{\hat{H}F}\right]\hat{H}\dot{\hat{\mathcal{R}}}_{k}+F\frac{k^{2}}{\hat{a}^{2}}\hat{\mathcal{R}}_{k}=0\,.
\end{equation}
In terms of the number of e-folds it can be written as 
\begin{equation}
\frac{\mathrm{d}^{2}\hat{\mathcal{R}}_{k}}{\mathrm{d}\hat{N}^{2}}-\left(3-\eh 1+\eh 2\right)\frac{\mathrm{d}\hat{\mathcal{R}}_{k}}{\mathrm{d}\hat{N}}+\frac{k^{2}}{\hat{a}^{2}\hat{H}^{2}}\hat{\mathcal{R}}_{k}=0\,,\label{R-EoM}
\end{equation}
where the Hubble flow parameters $\eh i$ are defined in eq.~(\ref{e1-def}).
The above equation is solved starting from Bunch-Davies vacuum initial
conditions 
\begin{equation}
\hat{u}_{k,\mathrm{vac}}=\frac{1}{\sqrt{2k}}e^{ik/(\hat{a}\hat{H})}\,,\label{eqa8}
\end{equation}
$\hat{N}=5$ e-folds before the horizon exit, which is defined as
$k=\hat{a}\hat{H}$, until $\hat{N}=5$ e-folds after the horizon
exit. The latter $5$ e-folds are added in order to make it certain
that the decaying mode is negligible and $\hat{\mathcal{R}}_{k}$
remains constant afterwards, whereas we start integrating $5$ e-folds
before horizon-crossing time to ensure that $k/(\hat{a}\hat{H})\gg1$
so eq.~(\ref{eqa8}) is a good approximation for the field perturbations.

The power spectrum of $\hat{\mathcal{R}}$ and the spectral tilt are
computed using 
\begin{eqnarray}
A_{\mathrm{s}} & = & \frac{k^{3}}{2\pi^{2}}\left|\hat{\mathcal{R}}_{k}\right|^{2}\,,
\end{eqnarray}
and 
\begin{eqnarray}
n_{\mathrm{s}}-1 & = & \frac{\mathrm{d}\ln A_{\mathrm{s}}}{\mathrm{d}\ln k}\,.
\end{eqnarray}

Similarly we compute the amplitude of tensor perturbations. In terms
of the number of $\hat{N}$ e-folds it is given by 
\begin{equation}
\frac{\mathrm{d}^{2}\hat{h}_{k}}{\mathrm{d}\hat{N}^{2}}-\left(3-\eh 1\right)\frac{\mathrm{d}\hat{h}_{k}}{\mathrm{d}\hat{N}}+\frac{k^{2}}{\hat{a}^{2}\hat{H}^{2}}\hat{h}_{k}=0\,,\label{h-EoM}
\end{equation}
with initial conditions 
\begin{equation}
\hat{h}_{k,\mathrm{vac}}=2\frac{\hat{u}_{k,\mathrm{vac}}}{\hat{a}}.\ 
\end{equation}

The tensor-to-scalar ratio is given by 
\begin{equation}
r\equiv\frac{A_{\mathrm{t}}}{A_{\mathrm{s}}}\,,
\end{equation}
where the tensor amplitude is defined by 
\begin{eqnarray}
A_{\mathrm{t}} & = & \frac{k^{3}}{2\pi^{2}}\left|\hat{h}_{k}\right|^{2}\,.
\end{eqnarray}

As we integrate eqs.~(\ref{num-EoM})-(\ref{num-dH}) and (\ref{R-EoM}),
(\ref{h-EoM}) we also compute the coordinate time, using eq.~(\ref{N-def-Ef})
\begin{equation}
t_{\textrm{end}}-t=\intop_{0}^{\hat{N}}\frac{\mathrm{d}\hat{N}}{\sqrt{F}\hat{H}}\,,
\end{equation}
where $t_{\textrm{end}}>t$ and $t_{\e}$ is an arbitrary value at
the end of inflation.

In the Jordan frame we only need to integrate homogeneous equations.
To make sure we start from the same spatial slice, the initial values
of $\phi$ and $\dot{\phi}$ are taken exactly the same as in the
simulations above. Then, using definitions of $\hat{N}$ and $N$
in eqs.~(\ref{N-def-Ef}) and (\ref{N-def-Jf}) together with eqs.~(\ref{N-F}),
(\ref{a-F}) we can relate the derivatives in both frames by 
\begin{equation}
\frac{\mathrm{d}\phi}{\mathrm{d}N}=\frac{1}{1+\frac{F_{,\phi}}{2F}\frac{\mathrm{d}\phi}{\mathrm{d}\hat{N}}}\frac{\mathrm{d}\phi}{\mathrm{d}\hat{N}}\,.
\end{equation}

The homogeneous system of Jordan frame equations that we are solving
are given in eqs.~(\ref{EoM-Jf})-(\ref{dH-Jf}). In terms of e-fold
numbers $N$ they can be written as 
\begin{align}
 & \frac{\textrm{d}^{2}\phi}{\textrm{d}N^{2}}-\left(3-\epsilon_{1}\right)\frac{\textrm{d}\phi}{\textrm{d}N}+\frac{V_{,\phi}}{H^{2}}=3F_{,\phi}\left(2-\epsilon_{1}\right),\\
 & H^{2}=\frac{V}{3F}\frac{1}{1-\frac{1}{6F}\left(\frac{\textrm{d}\phi}{\textrm{d}N}\right)^{2}+2\theta_{1}},\\
 & \epsilon_{1}=-\frac{\left(\frac{\textrm{d}\phi}{\textrm{d}N}\right)^{2}+\frac{\textrm{d}F}{\textrm{d}N}+\frac{\textrm{d}^{2}F}{\textrm{d}N^{2}}}{2F\left(1+\theta_{1}\right)},
\end{align}
where $\theta_{1}$ is defined in eq.~(\ref{theta1-def}). The coordinate
time is calculated by integrating (c.f. eq.~(\ref{N-def-Jf}))
\begin{equation}
t_{\textrm{end}}-t=\intop_{0}^{N}\frac{\mathrm{d}N}{H}\,.
\end{equation}

Having computed the time variable $t$ in both frames, we can compare
the results from the Jordan and Einstein frames on the same time slice.
The results are shown in figure~\ref{fig:comparison}. In that figure
we compare the exact Einstein frame calculations with the corresponding
Jordan frame slow-roll approximations. In particular, in figures~\ref{fig:e1}
and \ref{fig:e2} we see that the agreement between the first two
Einstein frame Hubble flow parameters and their expressions in terms
of Jordan frame flow parameters agree very well. While looking at
figures~\ref{fig:As}, \ref{fig:ns} and \ref{fig:r} we can conclude
the same about the exact simulations of inflation observables and
their approximate expressions in terms of Jordan frame flow parameters.

\begin{figure}
\begin{centering}
\includegraphics[scale=0.4]{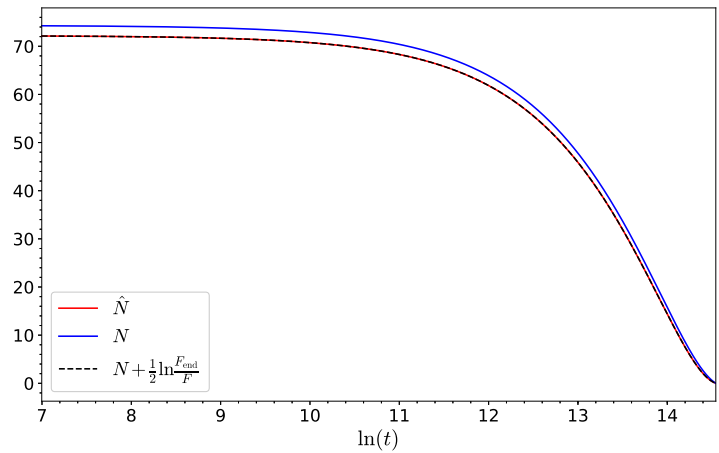}
\par\end{centering}
\caption{\label{fig:shift-e-folds}The e-fold number in each frame after solving
numerically the exact field equations. We show the effect of the e-fold
shift number as in eq.~(\ref{thN-SR-Jfrm}).}
\end{figure}

We can also check numerically the exact relation between $\hat{N}$
and $N$ in eq.~(\ref{thN-SR-Jfrm}). We plot their values on the
same spatial slice in figure~\ref{fig:shift-e-folds}. As one can
see, adding the e-fold shift number makes both curves, $\hat{N}\left(t\right)$
and $N\left(t\right)$, overlap exactly.

\begin{acknowledgments}
The work of M.K. and J.J.T.D. is partially supported by the Communidad
de Madrid \textquotedblleft Atracción de Talento investigador\textquotedblright{}
Grant No. 2017-T1/TIC-5305 and MICINN (Spain) project PID2019-107394GB-I00.
During the preparation of this work J.J.T.D. also received a grant
\textquotedblleft Ayudas de doctorado IPARCOS-UCM/2021\textquotedblright{}
from the Instituto de F\'{i}sica de Part\'{i}culas y del Cosmos
IPARCOS.
\end{acknowledgments}

\bibliographystyle{JHEP}
\bibliography{EJframes.bbl}

\end{document}